\documentclass[journal]{IEEEtran}
\usepackage[utf8]{inputenc}
\usepackage[english]{babel}
\usepackage{multicol} % Kolonner
\usepackage[a4paper, margin = 0.75in]{geometry} % Struktur

\usepackage{mathtools} % Matte
\usepackage{amsmath}
\usepackage{graphicx} % Grafikk
\usepackage{float}
\usepackage{fancyhdr} % Header

\usepackage{comment}
\usepackage[table]{xcolor}
\usepackage{multirow}

\newcommand{\stine}[1]{\textcolor{teal}{SFM: #1}}

\usepackage{caption}
\usepackage{subcaption}

\usepackage[ruled,vlined,linesnumbered]{algorithm2e}

\usepackage{hyperref}
\usepackage{xurl}
\usepackage{csquotes}
\usepackage[capitalize]{cleveref}
\usepackage{makecell}
\usepackage{multicol}

\usepackage{nomencl}
\makenomenclature

\title{A Study on V2G impact on the Reliability of Modern Distribution Networks} %V2G Support for Reliability of Electricity Supply of Modern Distribution Networks
\author{
\IEEEauthorblockN{Stine Fleischer Myhre\IEEEauthorrefmark{1}, Olav Bjarte Fosso\IEEEauthorrefmark{1}, Oddbjørn Gjerde\IEEEauthorrefmark{2}, Poul Einar Heegaard\IEEEauthorrefmark{3}} \\
\IEEEauthorblockA{\IEEEauthorrefmark{1}Department of Electric Power Engineering, NTNU\\
\IEEEauthorrefmark{2}SINTEF Energy Research\\
\IEEEauthorrefmark{3}Dept. of Information Security and Communication Technology, NTNU \\
Trondheim, Norway\\
\{stine.f.myhre, olav.fosso, poul.heegaard\}@ntnu.no; oddbjorn.gjerde@sintef.no}
}
\date{November 2021}

\begin{document}
\maketitle

\begin{abstract}
    \label{abstract}
        The increased penetration of electrical vehicles (EVs) in the distribution system creates a power system with many small moving batteries. With better charging technologies and market structures, the opportunity of using EVs for reliability purposes increases. By utilizing EVs for Vehicle-to-Grid (V2G), they can support the distribution network during outages or other issues. In this paper, we aim to investigate the impact V2G has on the reliability of modern distribution networks. We propose three new EV-oriented reliability indices to provide an overview of the impact experienced by the EVs in the system. For the reliability study, we use our developed reliability assessment tool for distribution networks. The tool is an open available software where the example network and datasets are embedded. We have extended this reliability tool to include a method for evaluating and calculating the reliability of distribution networks containing EVs with V2G possibilities. The paper also includes a thorough sensitivity analysis to further investigate the V2G impact of different parameters.

        %The electrical distribution system is moving towards a more decentralized, complex, and dynamic system. The system is experiencing a higher penetration of distributed energy resources, flexible resources, and active end-users, leading to an active distribution system. If these components are used actively, they can make a positive effect on the distribution system's reliability. 
        %This paper aims to investigate how a microgrid with renewable energy sources and energy storage might influence the reliability of electricity supply in a radially operated distribution system. The reliability is investigated from both the distribution system perspective and the microgrid perspective with the application of different scenarios. A reliability assessment method for modern distribution systems based on Monte Carlo simulations is presented. The method includes load flow calculations to capture the behavior of the system and system components. The model is tested on the IEEE 33-bus network. The result is confirmed through statistical testing showing the statistical significance in providing support from the microgrid on the distribution system's reliability. 
    
    \end{abstract}


\section{Introduction}
\label{sec:introduction}

The European Union (EU) has a goal to reduce greenhouse gas emissions by at least 55\% by 2030 compared to 1990 levels, and by 2050, become climate-neutral \cite{EU_target}. The greenhouse gas emissions from the transport sector constitute a large portion of the total greenhouse gas emissions in most countries. The transportation sector alone accounted for 24.6\% of the greenhouse gas emissions in the EU during 2018 \cite{share_EU}. In Norway, the greenhouse gas emission from the transportation sector %both road transport and other transport, 
was 15.6\% in 2020 \cite{share_Norway}. Due to the aim to reduce global greenhouse gas emissions, an increase in electrification is taking place. %In order to decreases the emissions from the transport sector, an electrification of the transport sector with green energy is necessary. 

%Multiple countries initiate measures to ban the sale of carbon emitting cars in the future. The Norwegian government aims to have all . the prognosis 
%With an increased penetration of electrical vehicles (EV) the potential of vehicle-to-grid (V2G) big. Using the cars as a support to the network. 

%There is an increased electrification with the mission to reduce the global greenhouse gas emissions. 
%With the Paris Agreement, 191 countries have joined a legally binding agreement on reducing the global greenhouse gas emissions to limit the global temperature to increase to two degrees Celsius. 

Several countries have initiated measures to ban the sale of new carbon-emitting passenger cars. %in the future to decrease the greenhouse gas emission from the transport sector. 
In the UK, the sale of petrol and diesel cars is planned to end by 2030 \cite{EV_UK}, the EU is planning to ban the sale by 2035 \cite{EV_EU}, while in Norway only zero-emission cars are to be sold from 2025 \cite{NTP_EN}. This will result in a higher share of electric vehicles (EV)s on the road. According to a study conducted by The Institute of Transport Economics, EVs will constitute on average 45.9--61.2\% of all cars by 2030 \cite{TOI}. However, this might differ between the regions of the country, resulting in regions with a high share of EVs and others with a low share of EVs. 

With a high share of EVs in the distribution systems, the potential of using EVs as grid support is increasing. By utilizing the EVs with, for example, vehicle-to-grid (V2G) technology, demand response measures such as frequency regulation, voltage support, and reactive and active power support can be possible. With V2G, the EVs that are connected to a charging station can support the network by transferring active and/or reactive power back to the power grid, becoming, in this way, small mobile batteries.

%This might result in places with a very high EV fleet, while others might have low.

%In Norway, the prognosis is that by 2030 the average share of EVs in the passenger car fleet will be in between 45.9 - 61.2\% \cite{TOI}.  

\subsection{Related work}
Some studies have investigated the potential of utilizing EVs with V2G to support the grid. V2G technology has been investigated for voltage regulation in distribution systems \cite{mazumder2020ev}, primary frequency regulation \cite{zecchino2019large, bayati2019short}, and regulation of small microgrids \cite{dinkhah2019v2g}. In \cite{yumiki2022autonomous}, a method for primary frequency regulation from an EV fleet is proposed. The study concludes with positive results and adds recommendations for ensuring a safe and stable operation when utilizing V2G. 

An interesting study on a military microgrid system is conducted in \cite{masrur2017military}. The study investigates V2G and vehicle-to-vehicle possibilities for power generation in a military microgrid. The study reports economic benefits due to reduced fuel costs and a benefit better than the already existing solution, indicating that EVs have an impact. However, the study points out the need for technology upgrades.   

Some other studies have investigated the possibility of increased reliability with V2G support. In \cite{farzin2016reliability}, a framework based on a non-sequential Monte Carlo simulation is developed to evaluate the reliability of a modern distribution system. The framework is used to investigate how integrated renewable generation and EV parking lots influence the reliability. The results indicate a significant improvement in which V2G is activated. A Sequential Monte Carlo Simulation (SMCS) framework is developed in \cite{xu2015reliability} where V2G and vehicle-to-home for reliability purposes are investigated.

The studies indicate a benefit of utilizing the available EVs for V2G services of the power system. However, the amount of support can be limited by multiple factors such as the EV availability and the technology related to charging, and will need to be further evaluated. In addition, in the literature, the impact on the degradation of the EVs is not considered to any great extent. 

%In \cite{9144526}, 

\subsection{Contribution}
In this paper, we aim to investigate the impact V2G service has on the reliability of modern distribution networks for short repair times. The paper utilizes V2G to increase the reliability of radially operated distribution networks. By providing three new EV-related indices, we aim to identify and investigate the impact experienced by the EVs when V2G is used. In addition, we provide a general methodology for evaluating how EVs perform from a reliability perspective through the use of RELSAD.
%This paper investigates how the utilization of V2G can help increase the reliability of electricity supply during failures in a radially operated distribution network. 
The study will focus on failures resulting in repair times of up to 2 hours, and the impact of V2G support will be investigated for the given down time. %The medium period outages in the system \textemdash outage periods of up to 2 hours. 

The contribution of this paper is to propose a reliability assessment method that considers V2G properties with a study to evaluate the impact on the reliability of modern distribution network. The method is an extension of our developed reliability assessment tool, RELSAD (RELiability tool for Smart and Active Distribution networks) \cite{Myhre2022}. RELSAD is an open-source software with an associated documentation page \cite{RELSAD_documentation}. In this paper, RELSAD is extended to include EVs and V2G opportunities. The main contributions of this research paper are:
\begin{itemize}
    \item A general methodology to assess the reliability of modern distribution networks with modeled V2G services. The potential and impact of V2G support for increased reliability of modern distribution systems will be assessed.
    \item Provide three new EV-related indices for describing the impact V2G services has on the EVs in the distribution network. 
    \item Extensive sensitivity analysis to provide a greater picture of the method and the impact of V2G support. The sensitivity analysis is conducted as a full factorial design. The evaluated parameters are repair time distribution, the charging capacity of the EVs, and the share of EVs in the network. 
    \item Provide a study on a realistic and modern distribution system. This is done systematically by gathering data for the system through procedures described in Sec \ref{sec:method} and Sec \ref{sec:casestudy}. This include using statistics from the Norwegian distribution systems, the Norwegian Electric Vehicle Association, and The Institute of Transport Economics in Norway. The proposed method is demonstrated through a case study on the IEEE 33-bus distribution network.  
\end{itemize}

\subsection{Paper structure}
The rest of this paper is organized as follows: Section \ref{sec:EV} introduces EVs. In Section \ref{sec:method}, the reliability assessment approach is presented along with how the EVs are implemented. The developed EV indices are outlined in Section \ref{sec:method} as well. The case study is discussed in \ref{sec:casestudy} before the results are presented and discussed in Section \ref{sec:result}. Lastly, some concluding remarks will be offered in Section \ref{sec:conclusion}. 

\begin{comment}

\begin{itemize}
\item Say something about the motivation for increased EV, that the countries are going to reduce their emissions.
\item Finne ut hvor mye transport står for av klimagassutslipp i Europa. 
\item Elektrifisering avincludesort
\item Snakke om Norge? siden vi har mest, nevne godene som også øker salg av elbiler? 
\item Snakke noe om hvordan nettet blir mer usikkert og trenger fleksibilitet
\item Sett det mot 2030 (kanskje 2050 rammeverk) 
\item Prediction of the amount of EV in the next ten? Years
\item Literature review of research that have investigated similar things
\end{itemize}

General notes: 
Norconsult raport:
Samfunnsøkonomisk lønnsomhet ligger til grunn
Klimagass i Norge ned 50 - 55\% innen 2030
Nullutslipsskjøretøy innen 2030 
Prognoser sier at 7 av 10 biler i 2030 er el i Norge
Hurtiglader kun på lange strekninger? 
Hjemmeladning 9 av 10 tilfeller
Reiser på mer enn 2 time lang reise trenger ladestasjon på veien
5-10\% har ikke ladestasjon hjemme 
Ladestasjoner bør plasseres spredt og med mange ladepunkt avgangen. 
Ladestasjon langs hovved vei 50 ladepunkter - trenger strømforsyning på 10 MakeLowercase
For lite ladere langs veien, trenger 1000 - 1200 nye ladestasjoner hvert år. 

except energy not charges? Should this be included? 

\end{comment}
\section{Theory}
\label{sec:EV}

This section explores the current practice of EV charging, battery degradation, and how EVs can be used for ancillary services in the network. 

%discuss theory related to the battery of EVs, charging of EVs, and the use of EVs to ancillary services. 

%\subsection{EVs in the future power system}

%Bruk resultater fra TØI artikkelen for å snakke litt om hvordan det blir i fremtiden med elbiler. 

\subsection{Charging infrastructure}

There exist multiple infrastructures for EV charging. The infrastructure can roughly be divided between home charging and charging stations with different charging performances such as fast charging and high-performance charging. The charging capacity at the charging stations can differ, from slow chargers with a capacity below 22 kW to super chargers with a capacity of over 100 kW \cite{IEA_EV}. Home charging, however, is a slower charging method, often with power capacities varying between 2-11 kW \cite{sorensen2021analysis}. The charging capacity is dependent on the car, the installed capacity of the charger, the AC or DC charger, and the voltage output of the network. There are two different methods for home charging, charging through charging boxes or through a socket. 

Charging methods and practices differ between countries \cite{IEA_EV}. The reason for this is the coverage of charging stations compared to EV owners with their own home charging. In Scandinavian countries, it is more common to have home charging. Public charging stations typically have higher charging capacities and are often used during short periods of time. However, in Asia, the share of public charging stations with slow charging is high.

\subsection{Battery degradation}

During the lifetime of an EV, the battery will experience degradation. The degradation of the battery is a result of multiple factors, such as structure degradation of the cells in the battery, charge, and discharge of the battery, temperature, and the aging and use of the EV \cite{thingvad2018influence}.  

In \cite{wang2016quantifying}, a methodology for quantifying the difference in battery degradation with and without V2G services is proposed. The paper provides a case study investigating how different support mechanisms, such as peak shaving, frequency regulation, and net load shaping, impact EV battery degradation compared to the cases where EVs do not offer grid services. The results illustrate that the battery degradation as a result of V2G services is minor, and the paper concludes that if the V2G services are only used in emergency events, the degradation will be inconsequential. The case study conducted in \cite{thingvad2018influence} supports this result. Here, battery degradation from providing primary frequency control support is investigated. The result illustrates that the battery degradation is 2\% when an EV supports the grid with 9 kW every day for five years.

Charging and discharging of a battery will over time contribute to the total degradation of an EV battery. Charging/discharging with a high rate results in higher battery degradation compared to slow charge/discharge. However, if the service is used rarely and the support is limited, the battery will not experience a large degradation as a result of V2G services. 

%a case study is conducted on investigating the EVs battery degradation for different factors including V2G support with primary frequency control. The study shows that the additional degradation of the battery for an EV that support the grid with 9 kW every day for 5 years is 2\%, and conclude that the degradation from the support is not very signifcant. 

\subsection{EVs used for ancillary services}

Charging boxes allow for smart home charging. Here, the opportunities for an EV owner to activate different charging strategies can be made possible. Through such a system, the EV owner could, for example, decide to be a part of a flexibility market. In such a flexibility market, the EV owner could agree to let a power supplier use the EV for V2G service, and in return be compensated. 

In Ref. \cite{statnettibber}, a pilot project is initiated to investigate how different technologies including EVs are used for providing flexibility in an electronic bid ordering market. The EV owner could choose to be a part of the flexibility market through their smart home charger controlled by the supporting electricity company. 
%A pilot project between the Norwegian Transmission System Operator (Statnett), the electricity supplier Tibber, and the energy provider Entelios has been conducted on increasing the availability of end-costumer flexibility \cite{statnettibber}. In the study, EVs are among the technologies used for providing flexibility in an electronic bid ordering market. The EV owner could choose to be a part of the flexibility market through their smart home charger controlled by Tibber. 
During peak load hours, the EVs participating could be used to decrease the load in the power system by pausing the charging of the EVs. %Companies such as Tibber, Plexigrid, and Virta develop platforms for smart controlling of technologies such as EVs. 
In the future with new standards and business models, V2G can be a part of the market solutions and constitute an important role in flexibility and reliability in the power system \cite{Virta}.

\section{Methodology}
\label{sec:method}

This section aims to present the developed reliability method and the proposed EV-oriented reliability indices. In this section, we will first present the implemented reliability indices with the proposed EV-oriented reliability indices. Second, we outline the approach for the used reliability attributes in the system. Third, the methodology used for estimating the availability of EVs is described before the repair time distribution of the components is given. In the end, RELSAD, the developed reliability assessment tool, is presented and the implementation of V2G is discussed. 

%First, we present the implemented load flow method and optimization problem. Second, we present both common reliability indices and developed indices for EVs. Third, the methodology used for estimating the availability of EVs and the outage time of components in the network are presented. In the end, the functionality of RELSAD with the extension developed in this paper for including EVs and V2G possibilities are presented. 

%Include the math for battery charge and discharge 

%V2G mode of the battery. 

%Talk about the distributions for both number of EVs and the SoC level and the outage time. 
%In this section, we present the developed reliability methodology for networks including EVs and V2G. First, we present the 
%Then, the implemented ... is presetend. Finally, we present the developed reliability assessment method ... 
\subsection{Reliability indices}
\label{sec:rel_indices}

There are different reliability indices used for measuring and quantifying the reliability in distribution networks. They can be divided into \textit{customer-oriented indices} and \textit{load- and production-oriented indices} \cite{Billinton}. In this paper, three classically used reliability indices are included. In addition, we have developed three indices related to the impact the EVs experience during V2G operation.

The three basic reliability parameters used for calculating reliability indices are: 1) the fault frequency or the average failure rate, $\lambda_s$, 2) the annual average outage time, $U_s$, and 3) the average outage time, $r_s$. 

\begin{comment}
In a radial network, the three different reliability parameters can be calculated as in eq. \ref{eq:faultfreq}, eq. \ref{eq:avgoutagetime}, and eq. \ref{eq:outagetime}. Here, $\lambda_i$ and $r_i$ are the failure rate and outage time at load point $i$, respectively. These equations are used to state the load-point reliability. 
%However, these do not tell anything about the electrical consequence of a failure or the cost related to a failure.
 
\begin{equation}
    \label{eq:faultfreq}
    \lambda_{s} = \sum \lambda_{i}
\end{equation} 

\begin{equation}
    \label{eq:avgoutagetime}
    U_{s} = \sum \lambda_{i} r_{i}
\end{equation}

\begin{equation}
    \label{eq:outagetime}
    r_{s} = \frac{U_{s}}{\lambda_{s}} = \frac{\sum \lambda_i r_i}{\sum \lambda_i}
\end{equation}

\end{comment}

\subsubsection{Load- and Production-oriented indices}
The load- and production-oriented indices aim to indicate the electrical consequence of faults in the system \cite{Billinton}. In this paper, the total Energy not Supplied (ENS) in a system is investigated. ENS can be calculated as:

\begin{equation}
    \label{eq:ENS}
    {\tt ENS}_{s} = U_{s}P_{s}
\end{equation}

\begin{comment}
The interruption cost for the system can be calculated as seen in eq. \ref{eq:cost} \cite{943073}.
Here, $c_{i}$ is the specific interruption cost for each customer category at load point $i$. 

\begin{equation}
    \label{eq:cost}
    {\tt CENS}_{s} = \sum {\tt ENS}_{i}c_{i}
\end{equation}

\end{comment}

\subsubsection{Customer-oriented indices}

The customer-oriented indices aim to indicate the reliability of the distribution system based on the interruption experienced by the customers \cite{Billinton}. In this paper, three important indices will be investigated:

\begin{enumerate}
    \item System Average Interruption Frequency Index% (SAIFI)
\begin{equation}
    \label{eq:SAIFI}
    {\tt SAIFI} = \frac{\sum_{\forall i} \lambda_{i}N_{i}}{\sum N_{i}}
\end{equation}
where $N_{i}$ is the {\em total number of customers served} at bus $i$, and $\sum_{\forall i} \lambda_{i}N_{i}$ is the {\em total number of customer interruptions}. ${\tt SAIFI}$ is a measure of the frequency of interruptions the customers in the system expect to experience. Any interruption seen by the consumer is counted as a fault, regardless of origin. 
\newline
\item System Average Interruption Duration Index% (SAIDI),
\begin{equation}
    \label{eq:SAIDI}
    {\tt SAIDI} = \frac{\sum U_{i}N_{i}}{\sum N_{i}}
\end{equation}
where $\sum_{\forall i} U_{i}N_{i}$ is {\em total number of customer interruption} durations. ${\tt SAIDI}$ is a measure of the expected duration of interruptions a customer is expected to experience. 

\begin{comment}
\newline
\item Customer Average Interruption Duration Index% (CAIDI). 
\begin{equation}
    \label{eq:CAIDI}
    {\tt CAIDI} = \frac{\sum U_{i}N_{i}}{\sum_{\forall i} \lambda_i N_{i}} = \frac{\tt SAIDI}{\tt SAIFI}
\end{equation}
${\tt CAIDI}$ is the ratio between ${\tt SAIDI}$ and ${\tt SAIFI}$ and measures the average duration each given customer in the system is expected to experience. 
\end{comment}
\end{enumerate}

\subsubsection{EV-oriented indices}
Since the EVs will be affected by outages in the distribution system, their own EV indices are established. 

\begin{enumerate}

    \item Average EV Demand Not Served

    \begin{equation}
        \label{eq:EVAENS1}
        {\tt EV_{Demand}} = \sum U_{EV_i}(P_{ch_i}+P_{dis_i})
    \end{equation}
    where, $U_{EV_{i}}$ is the expected outage time EV $i$ experience, $P_{ch_{i}}$ is the amount EV $i$ charges or wants to charge, while $P_{dis_{i}}$ is the amount of power EV $i$ discharges. This index aims to give an indication of how much power the EVs are not being served. During outages in the system, an EV can experience periods where the EV is unable to charge as a result of the ongoing fault in the system. When an EV is used for V2G services, the index will increase since the EV is discharging power in addition to having a demand. This means that the amount of discharged power from the EV will be added to the demand not served for the EV. This index can be used to give an indication of how much power an EV contributes and the amount of demand from the EVs in the system.

    \item Average EV Interruption Frequency
    \begin{equation}
        \label{eq:EVAIDI}
        {\tt EV_{Int}} = \frac{\sum_{\forall EV} \rho_{EV_{i}}N_{EV_{i}}}{\sum N_{EV_{i}}}
    \end{equation}
    Here, $\rho_{EV_{i}}$ is the expected amount of times an EV is used for V2G services,  $\sum N_{EV_{i}}$ is the total number of EVs in the distribution system, and $N_{EV_{i}}$ is the total number of EVs at a bus.
    This index is based on $\tt SAIFI$. 
    This index gives the average number of interruptions an EV experiences due to V2G services in the system. In a simulation, the fraction of EVs that are used for V2G services in an EV park is found and multiplied by the amount of EVs in the EV park to give an estimation of how many cars in that EV park are used for V2G. This index can be used to illustrate how often an EV is estimated to be used for V2G services. The index is sensitive to the frequency of failures in the system. With a higher frequency of failures, the system will require the EVs to support more frequently.

    \item Average EV Interruption Duration
    \begin{equation}
        \label{eq:EVAIFI}
        {\tt EV_{Dur}} = \frac{\sum U_{EV_{i}}N_{EV_{i}}}{\sum N_{EV_{i}}}
    \end{equation}
    This index is based on $\tt SAIDI$.
    The index gives the average duration of the EVs used for V2G services in the system. This index accumulates the amount of time every EV in the system is being used for V2G services and divides it by the number of EVs in the system. The index gives an indication of how long an EV can be expected to be used for V2G services during a given period. The index will increase with a longer down time of the system since the EVs will most likely be used for longer periods at a time.

\end{enumerate}

\subsection{Availability distribution of the EVs in the distribution network}
\label{sec:EV_availability}

For activating V2G support in the distribution network, the availability of the EVs in the network needs to be established. The availability of the EVs can be decided based on the number of EVs that are charging. Therefore, the availability can be estimated based on the charging patterns of EVs.  

\begin{figure}
    \centering
    \includegraphics[width=0.5\textwidth]{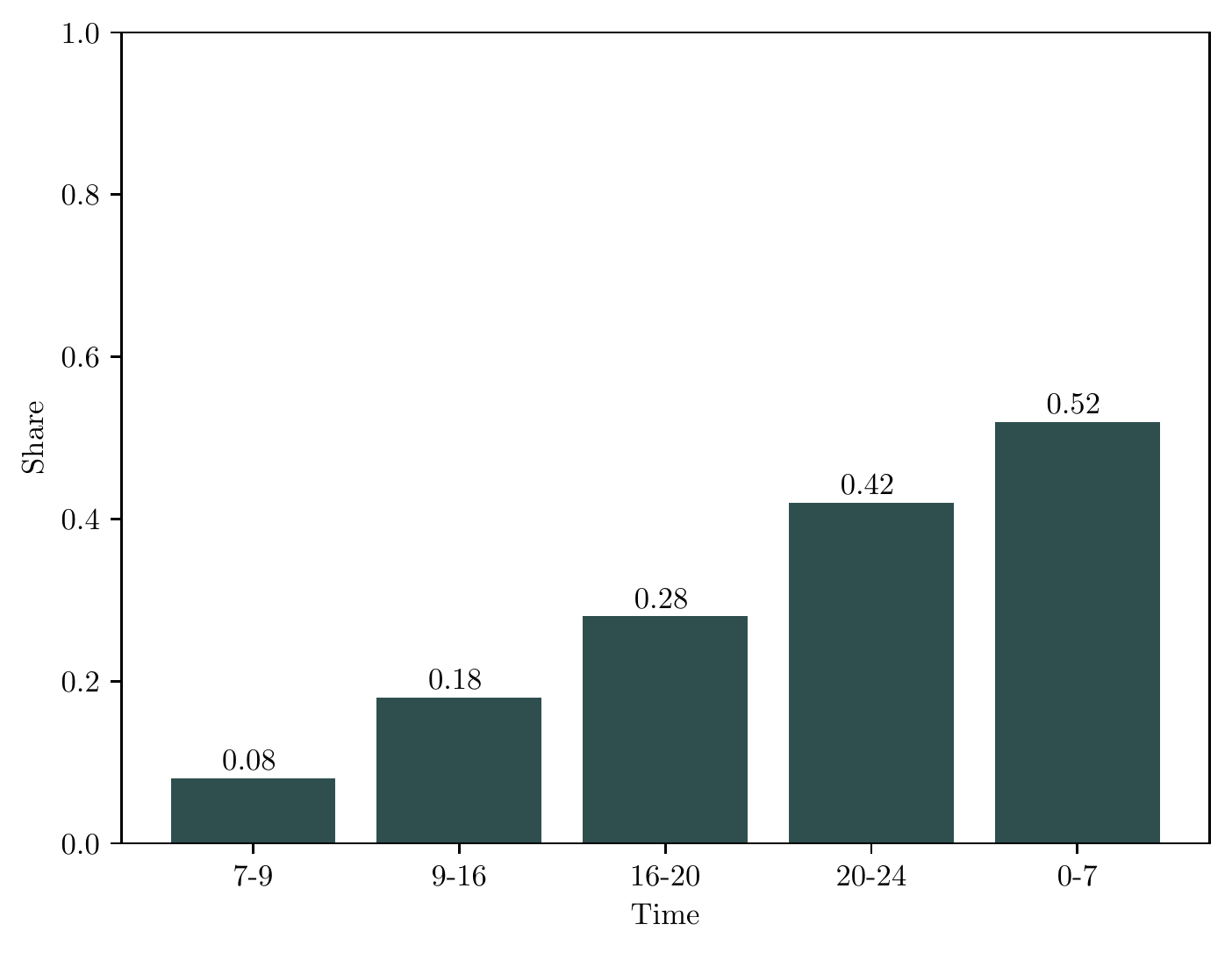}
    \caption{The normal charging time of an EV, based on \cite{elbilisten}.}
  \label{fig:Charging_time}
\end{figure}

\begin{figure}
    \centering
    \includegraphics[width=0.5\textwidth]{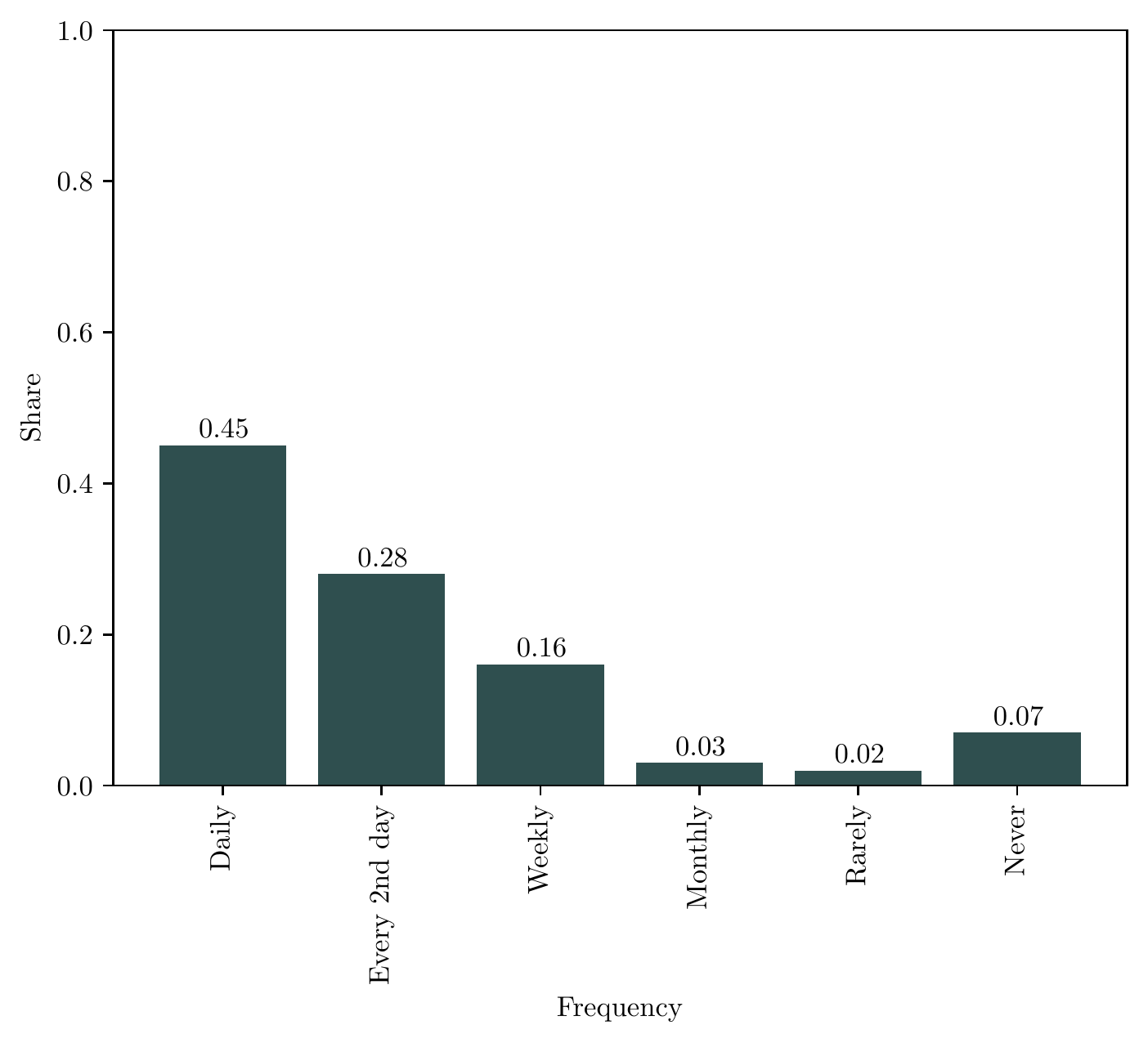}
    \caption{The home charging frequency of an EV owner, based on \cite{elbilisten}.}
  \label{fig:Charging_pattern}
\end{figure}

The charging pattern of Norwegian EVs is examined in a survey conducted by The Norwegian Electric Vehicle Association (\textit{Elbilforeningen}). The survey was conducted in 2019 with more than 16,000 respondents \cite{elbilisten} and will be used as a basis for estimating the availability of EVs in the distribution system. The result of the survey is summarized in Fig. \ref{fig:Charging_time} and \ref{fig:Charging_pattern}. Fig. \ref{fig:Charging_time} illustrates at which time the EV owners usually charge their EV. In \ref{fig:Charging_pattern}, the frequency for home charging is shown. These results correspond well with the result on charging patterns obtained in \cite{sadeghianpourhamami2018quantitive} and give a good estimation for evaluating the expected availability of EVs that are charging at home during a specific time in the distribution networks.

%Therefore, the survey is used as a basis for estimating the availability of the EVs in the distribution system. 

%\textit{Elbilforeningen} is a Norwegian electric vehicle association which represents Norwegian EV owners. In 2019, they conducted a survey on charging pattern of Norwegian EV owners with more than 16 000 respondents \cite{elbilisten}. The result on charging pattern from the survey correspond well with the patterns obtained in \cite{sadeghianpourhamami2018quantitive}.  

%It is predicted that the amount of EVs in the distribution system 
%The result from the survey is used to make a distribution of the availability of the EVs in the distribution system in this paper. 
Based on the distributions in Fig. \ref{fig:Charging_time} and Fig. \ref{fig:Charging_pattern} and the amount of EVs in the distribution system, the probability of an EV being charged at home is:

\begin{equation}
    \label{eq:EV_availability}
    A_{EV} = n_{customers}\cdot X_{EV}\cdot C(t)\cdot D_{EV}
\end{equation}

Here, $n_{customers}$ is the number of households in the distribution system, $X_{EV}$ is the percentage share of vehicles that are EVs in the distribution network, $C(t)$ is the share of EVs charging at time $t$, and $D_{EV}$ is the estimated daily charge frequency. The daily charge frequency is based on the result from Fig. \ref{fig:Charging_pattern} and gives a probability of the EV being charged at home. Based on Fig. \ref{fig:Charging_pattern}, $D_{EV} = 0.45+0.28\cdot0.5+0.16\cdot1/7+0.03\cdot1/30 = 0.61$.

\subsection{Repair time distribution}
\label{sec:outage_time_dist}

In this paper, events resulting in varied down times are considered. Therefore, the repair time of the system components will vary within a range of possible repair times. The repair time used in this paper is based on yearly reliability statistics for the Norwegian DSOs and is described in Sec \ref{sec:casestudy}. There exist multiple possible distributions that can be used to decide the repair time of an event, such as Log-normal-, gamma-, and normal distributions \cite{billinton_reliability_1994}. In this paper, truncated normal distributions will be used. 

%a range span of possible outage times that results in outages in the distribution system are investigated. There are multiple possible distributions that can be used to decide the outage time for a component such as Log-normal-, gamma-, and normal distributions \cite{billinton_reliability_1994}. In this paper, normal distribution will be used. The probability density function of a normal distribution is 

\subsubsection{Truncated normal distribution}

A truncated normal distribution has a probability distribution similar to a normal distribution but is bounded by either an upper or lower limit or both. 

\begin{figure}
    \centering
    \includegraphics[width=0.5\textwidth]{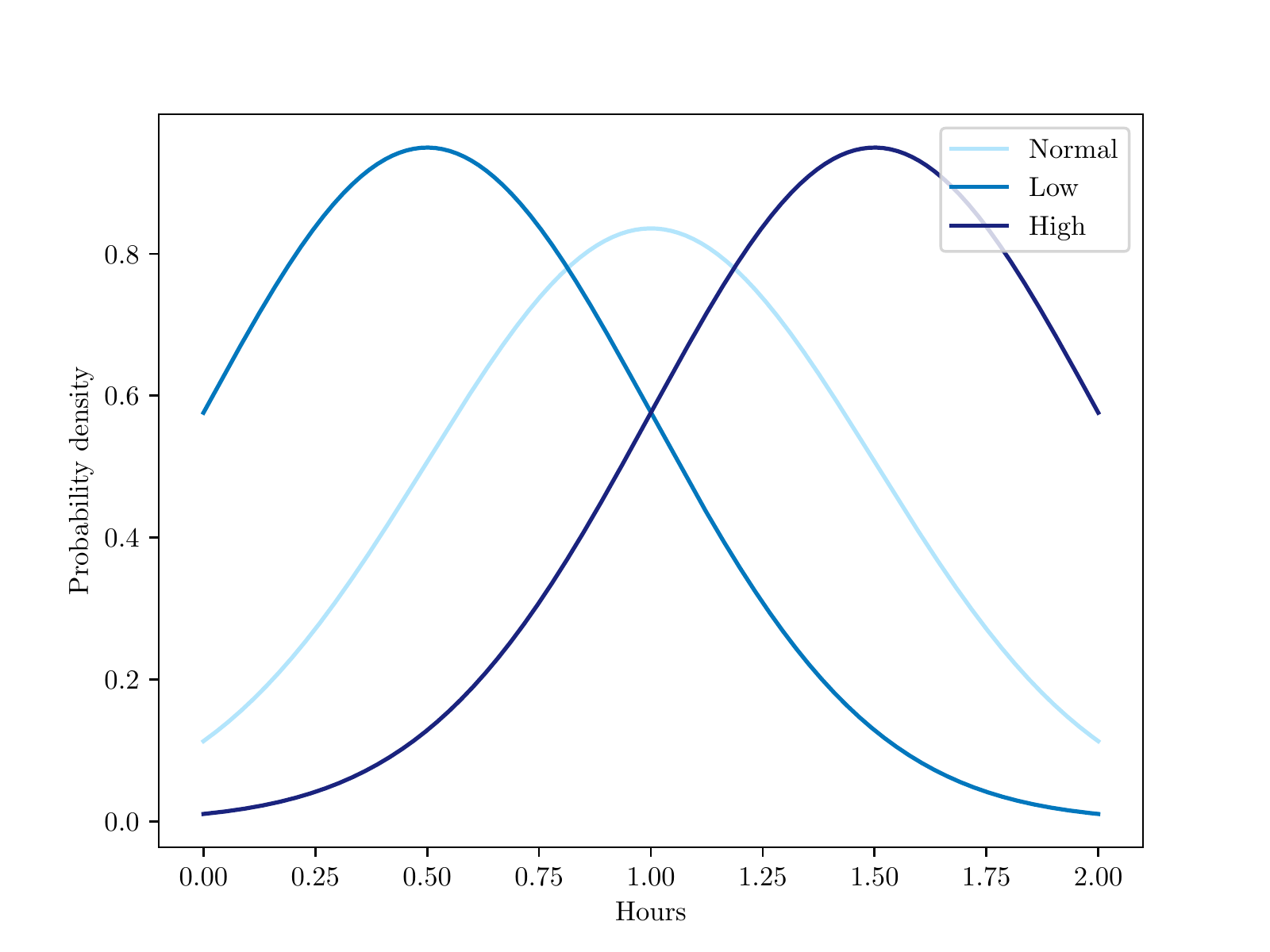}
    \caption{Truncated normal distributions of the repair time for three different location parameters (low, high, and normal).}
  \label{fig:outage_time_dist}
\end{figure}

The probability density function of a normal distribution can be expressed as

\begin{equation}
    \label{eq:NormalDistribution}
    f(x) = \frac{e^{-\frac{(x-\mu)^{2}}{(2\sigma)^{2}}}}{\sigma \sqrt{2\pi}}  
\end{equation}

where, $\mu$ is the location or mean parameter of the distribution and $\sigma$ is the scale parameter or standard deviation. 
Truncated normal distributions are good distributions to use for repair times that are limited by boundaries. Fig. \ref{fig:outage_time_dist} illustrates how the truncated normal distributions look with different location parameters (high, low, and normal mean).

\begin{comment}

\subsubsection{Gamma distribution}
For skewed location parameters, Gamma distribution is a more fitting distribution.  
The probability density function for the Gamma distribution can be expressed as

\begin{equation}
    \label{eq:GammaDistribution}
    f(x) = \frac{(\frac{x-\mu}{\beta})^{\gamma - 1} e^{(-\frac{x-\mu}{\beta})}}{\beta \Gamma(\gamma)}
\end{equation}

where $\gamma$ is the shape parameter, $\beta$ is the scale parameter, and $\Gamma$ is the function expressed as

\begin{equation}
    \label{eq:GammaFunction}
    \Gamma(a) = \int_{0}^{\infty} t^{a-1}e^{-t} dt
\end{equation}

For an outage time that can vary in a range between $r \in [r_{min}, r_{max}]$ where knowledge about the location and scale parameter of the possible outage distribution is lacking, three different distributions for the outage time can be made. 

Fig. \stine{Legge til plot av de tre mulige distribusjonene for utetid} illustrates three possible outage distributions. \stine{Legge til litt beskrivelse når plottet er på plass}

\end{comment}

%without the knowledge of the location and the scale of the distribution, 

\subsection{Reliability evaluation of modern distribution system - RELSAD}

The electrical power system is rapidly modernizing with higher penetration of renewable energy sources, flexible energy sources, and Information and Communication Technology (ICT). The modernization of the power system results in a more complex system structure with increased dependencies. The traditional methods for reliability assessment of power systems do not consider these changes. In response, we developed RELSAD---RELiability tool for Smart and Active Distribution networks. 

RELSAD is a reliability assessment tool for modern distribution networks based on SMCS. The software is developed as an open-source Python package and is published as a scientific tool \cite{Myhre2022}. Additionally, a documentation page for the software has been written \cite{RELSAD_documentation}. 
RELSAD was developed to give a foundation for the reliability assessment of modern distribution systems where the changed behavior and the increased complexity in the distribution network are considered. 

RELSAD provides the following features: 

\begin{itemize}
    \item Calculation of detailed results in the form of distributions and statistics on important reliability indices.
    \item Opportunity for investigation of network sensitivities where different parameters such as repair time, failure rate, load and generation profiles, placement of sources, and component capacities and their impact on the system.
    \item Implementation and investigation of diverse and advanced network constellations spanning from small passive distribution systems to large active networks including DGs, batteries, microgrids, and EVs where the networks are operated radially.
    \item The ability to investigate and analyze the reliability of Cyber-Physical Networks.
\end{itemize}

In this paper, RELSAD is extended to include EVs and V2G possibilities. This paper will describe the core functionality of RELSAD in addition to the contribution and implementation of EVs with V2G.

\subsection{Structure of RELSAD}

\begin{figure}
    \centering
    \includegraphics[width=0.5\textwidth]{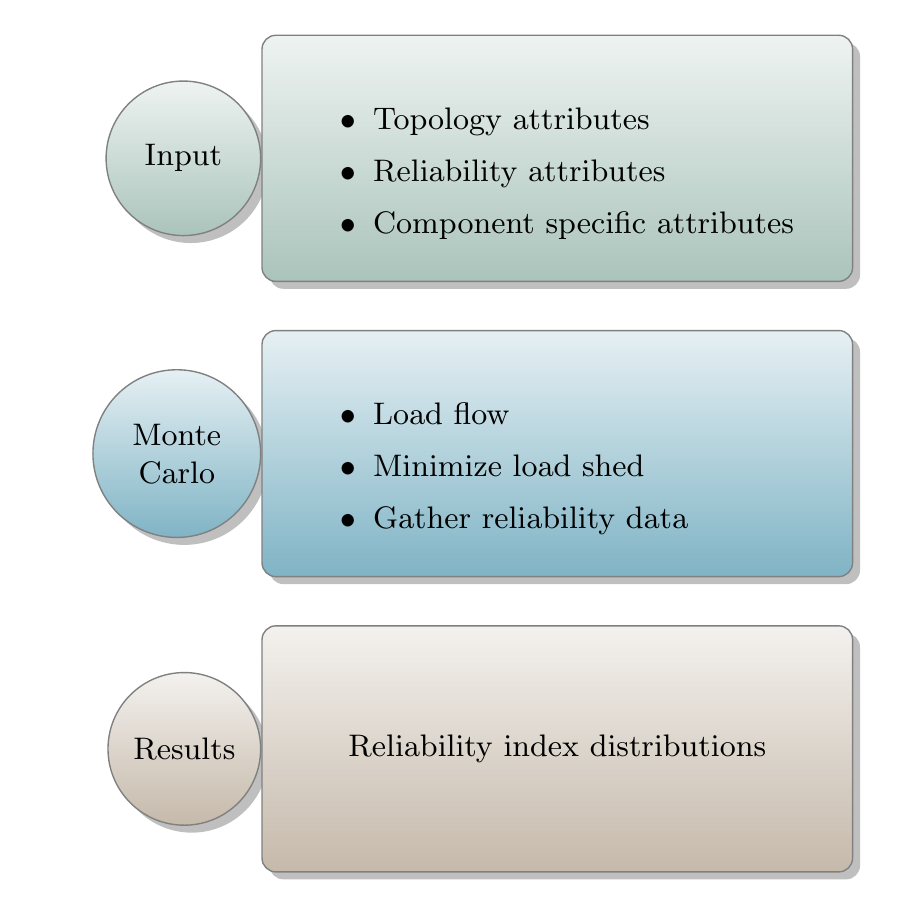}
    \caption{Overview of the RELSAD model. A full description can be found in the software documentation \cite{RELSAD_documentation}.}
  \label{fig:RELSAD_structure}
\end{figure}

In Fig. \ref{fig:RELSAD_structure}, an overview of the RELSAD model is illustrated. RELSAD considers different attributes such as 

\begin{itemize}
    \item \textbf{Topology attributes:} such as spatial coordinates and connected components.
    \item \textbf{Reliability attributes:} such as failure rates and repair time distributions.
    \item \textbf{Component specific attributes:} such as, for an EV, this could be the battery capacity, the inverter capacity, efficiency, and charging limitations. 
\end{itemize}

The core of the software is the simulation where different operations are performed. The general functionality of RELSAD is described in Sec. \ref{sec:core_func_RELSAD} and the EV and V2G specific functionalities are described in Sec. \ref{sec:V2G_implementation}. 

The output or results of the RELSAD software are given as reliability index distributions. Some of the different distributions that can be investigated are described in Sec. \ref{sec:rel_indices} such as the proposed EV-oriented reliability indices. 

RELSAD is constructed in an object-oriented fashion, where the systems and system components are implemented with specific features simulating their real-life behavior. RELSAD is a general tool that can consider any network that operates radially. In RELSAD, a power system - $P_{s}$ is first created before distribution systems - $D_{s}$ and other systems such as Microgrids - $M_{s}$ are created inside the $P_{s}$. After this, the electrical system components can be created and added to associated network layers. The possible electrical components are lines - $l$, buses - $b$, disconnectors - $d$, circuit breakers - $cb$, generation units - $p_{u}$, batteries - $p_{b}$, and EV parks - $p_{EV}$. In addition, a load - $p_{d}$ can be assigned to each bus. This is described in more detail in the documentation page of the software \cite{RELSAD_documentation}.

\subsection{Core functionality of RELSAD}
\label{sec:core_func_RELSAD}

In this section, the core functionality of RELSAD is described. The reliability assessment of a system is solved through SMCS in RELSAD. To account for the active participation of different technologies such as DGs, batteries, and EVs, a load flow solver is implemented to evaluate the electrical consequence of faults in the system. Through load flow calculations, the behavior of the system during different scenarios can be evaluated. In RELSAD, a Forward-Backward sweep (FBS) approach is implemented as the load flow solver. 
In the FBS approach, the load flow is calculated by updating the power flow through a backward sweep before the voltage magnitudes and angles at the system buses are updated in a forward sweep \cite{haque1996load}. 
%In RELSAD, the complexity of modern distribution systems is addressed and implemented. RELSAD is developed as an open-source Python package, available at \cite{RELSAD}. 
%Since the network is radially operated, a Forward-Backward sweep (FBS) approach is implemented as the load flow solver \cite{haque1996load}. 

Due to the potential island operation of parts of the distribution system, power balance needs to be ensured. To achieve this, a simple load shed optimization problem is included. The objective of the load shed optimization problem, seen in eq. \ref{eq:min}, is to minimize the total load shed in the network based on the price of shedding different load types. The price is based on the \textit{Cost of energy not supplied} (CENS). This is subjected to load flow balance and the capacity limitations over the power lines, the load, and the generation in the distribution system: 

\begin{align}
    &\underset{P^{s}_{n}}{\text{minimize}}
    \quad \mathcal{P}_s = \sum_{n = 1}^{N_n}   C_{n}\cdot P^{s}_{n} \label{eq:min}\\
    &\text{subject to: } \nonumber \\
    &\begin{aligned}
        \sum_{i=1}^{N_l} \alpha_i \cdot P^{l}_{i} &= \sum_{j=1}^{N_g} \nu_j \cdot P^{g}_{j} - \sum_{k = 1}^{N_n}& \eta_k& \cdot (P^{d}_{k} - P^{s}_{k})\\
        \min P_{j}^{g} &\leq P_{j}^{g} \leq \max P_{j}^{g} &\forall j&=1,\dots,N_{g}\\
        0 &\leq P_{k}^{s} \leq P_{k}^{d}  &\forall k&=1,\dots,N_{n}\\
        \left| P_{i}^{l} \right| &\leq \max P_{i}^{l}  &\forall i&=1,\dots,N_{l}\\
    \end{aligned} \nonumber
\end{align}

Here $C_n$ is the cost of shedding load at node n while $P_{n}^{s}$ is the amount of power shed at node n. $P_{j}^{g}$ is the production from generator $j$. $P_{k}^{d}$ is the load demand at node $k$ while $P_{i}^{l}$ is the power transferred over line $i$. $\alpha_i$ = 1 if line $i$ is the starting point, -1 if line $i$ is the ending point. $\nu_j$ = 1 if there is a production unit at node $j$, otherwise it is 0. $\eta_k$ = 1 if there is a load on node $k$, otherwise it is 0.

\subsubsection{Incremental procedure of fault handling in RELSAD}

The SMCS in RELSAD is constructed to have a user-chosen increment. This can, for example, be a second, a minute, an hour, or a day. 
The incremental procedure of the fault handling model implemented in RELSAD is illustrated in Algorithm \ref{alg:procedure}. After a $P_{s}$ is created with associated systems and components, the incremental procedure will set the load and generation at the system buses for the current time increment. Then the failure status of each component is calculated based on random sampling and the failure rate of the different components. If any component is in a failed state, the network with the failed components will be divided into sub-systems, and a load flow and load shedding optimization problem will be solved for each sub-system. The load flow and load shedding optimization problem are included to calculate the electrical consequence of a fault when active components are present. When this is performed, the historical variables of the components are updated. The historical variables can then further be used to evaluate the reliability of the $P_{s}$, the different networks in the power system, and the individual load points. An example of this is illustrated in the documentation page for the software \cite{RELSAD_documentation}.  
The rest of this section describes the different procedures of the reliability evaluation in greater detail. 

%First, the system is initialized and the parameters set. Then the algorithm will check for failures in the system. If there are any failures, the system will be divided into sub-systems and a load flow and minimization problem will be solved for each sub-system. 
%This paper does not consider a true Monte Carlo simulation since it uses the element of time and will therefore not simulate a fully random process. 

\begin{algorithm}
    \SetAlgoLined
     Set bus $p_d$ and $p_u$ for the current time increment\;
     Draw component fail status\;
     \If{Failure in $P_s$ }{
         Find sub-systems\;
         \ForEach{sub-system}{
            Update $p_b$ and $p_{EV}$ demand (charge or discharge rate, discharge of EVs if V2G is activated)\;
            Run load flow\;
            Run load shedding optimization problem\;
         }
         Update history variables\;
     }
    \caption{Increment procedure}
    \label{alg:procedure}
\end{algorithm}

\subsection{V2G implementation in RELSAD}
\label{sec:V2G_implementation}

EVs and V2G possibilities are implemented in RELSAD. The EVs are implemented to be connected to the buses in the system and as susch form an EV park at the bus. Furthermore, the number of EVs in an EV park can be decided based on a user-defined availability distribution. Each EV park can be flagged to allow for V2G possibilities or not. The specifications, such as battery capacity, inverter capacity, and min and max SoC, of the EVs in the EV park, can be set by the user. 

The behavior of the EVs in an EV park works similarly to that of a battery. The EVs can be charged and if V2G is possible, the EVs can be discharged. The charging and discharging of the EVs are restricted by the specifications of the EV.  

\subsubsection{V2G and EV algorithm}

In Algorithm \ref{alg:procedure_EV}, the EV park procedure during faults in the network is displayed. Since RELSAD is a reliability assessment tool aiming to evaluate the reliability of a system, only situations with faults are considered. It is presumed that when no faults in the network are present, the network is working correctly. 

The algorithm is called upon when a failure occurs in network $N$, in the current time increment. Then the number of cars in an EV park is determined based on an availability distribution. Next, the SoC state of each EV in the EV park is calculated following a given distribution (in this study, it follows a uniform distribution). The SoC state is stated between the maximum and minimum allowed SoC state. Then the network balance of network $N$, $N_b$ is calculated. This is in order to find out the demand of each EV and the load and generation of the network. 

If V2G is activated for the EV park, the EV will either be charged or discharged depending the load balance. If there is an available generation, the EVs can be charged; if there is a lack of generation, the EVs might be discharged to be used for support. After this, the load balance of the network is updated. 

If V2G is not activated, charging of the EVs is only possible if there is a surplus of generation in the network (or connected to the overlying network). The load balance of the network is then updated. 

In the end, the history variables of the EVs in the EV parks are updated.

\begin{algorithm}
    \SetAlgoLined
    \If{Failure in network, $N$, happened in current increment}{
        Draw number of cars\;
        \ForEach{car}{
            Draw SoC state\;
        }
    }
    Get network load balance, $N_{b}$\;
    \ForEach{car}{
        \eIf{Vehicle to grid mode is active}{
            Charge/discharge car battery\;
            Update $N_{b}$\;
        } {
            Charge car battery\;
            Update $N_{b}$\;
        }
    }
    Update history variables\;
    \caption{EV park procedure during faults in the network}
    \label{alg:procedure_EV}
\end{algorithm}

\begin{comment}
1. Må legge inn EV ladepark - denne skal ha plass til X antall biler og er av en satt størrelse med inverterkapasitet. 
2. Tilgjengeligheten av biler kan trekkes som en uniformfordeling
3. Tilgjengeligheten på batteriet kan trekkes som en uniformfordeling mellom min SoC og maks SoC. 
4. Samler opp det som blir trukket av antall biler og tilgjengelig kapasitet og legger til i SoCen til bilparken som bestemmer nivået av effekt som kan gis til nettet. 

\end{comment}

\subsubsection{Aggregated battery solution}

In this paper, the amount of EVs in the distribution network will be based on the number of households in the network. Therefore, EVs charged in places other than at home will not contribute to the study. The amount of EVs that in theory are able to contribute is decided based on the total amount of households in the distribution network. From there, predictions about the share of households that owns an EV can be decided.%This amount will make up the EV park that can be used in the study for flexibility support in the network. 

For simplicity, the model is built up to make an aggregated battery solution as seen in Fig. \ref{fig:anumberted_battery}. The aggregated battery idea illustrates the situation on a bus in the system. The bus has a given number of households, and based on the availability of EVs, measured as in Sec. \ref{sec:EV_availability}, a given number of EVs will be connected to the bus. From this, the SoC level of each EV will be estimated based on a uniform distribution between the minimum and maximum levels of the SoC in the EV. 
This information will then be added to the aggregated battery for the bus. The aggregated battery for the bus will contain information about the EVs on the bus, the aggregated amount of power in the battery, and the total charging capacity of the battery. 

%illustrating the EV situation of the bus including the total availability of power from the battery. The charging capacity of each home charger will also be aggregated. 

%Fig. \ref{fig:aggregated_battery} illustrates the aggregated battery idea. Here, a bus in the distribution system has X available EVs, which again each has their own SoC level in the EV battery. This information will then be added into an aggregated battery for each bus in the system where there are EVs available. The aggregated battery on the buses can change from time to time based on the number of available EVs at the bus and the SoC situation of each EV. This information decides how much each bus potentially can contribute during the failure in the system. 

\begin{figure}
    \centering
    \includegraphics[width=0.5\textwidth]{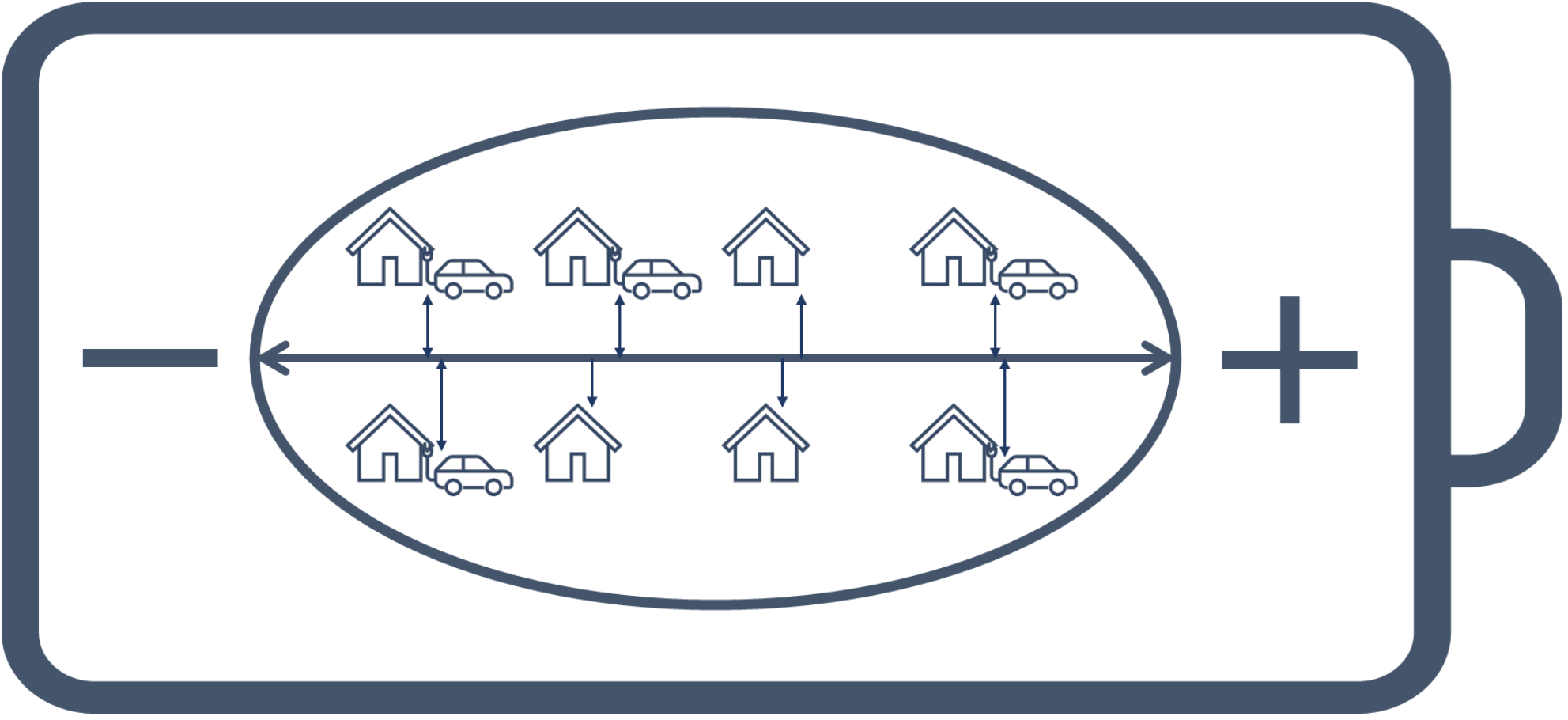}
    \caption{Aggregated battery solution for EVs at a bus in the system. The figure illustrates how an aggregated battery solution is created for an EV park for the EVs at a bus in the network.}
  \label{fig:aggregated_battery}
\end{figure}

\section{Case study}
\label{sec:casestudy}

\subsection{Procedures for developing the test network}

\begin{figure*}[tb]
    \centering
    \includegraphics[width=\textwidth]{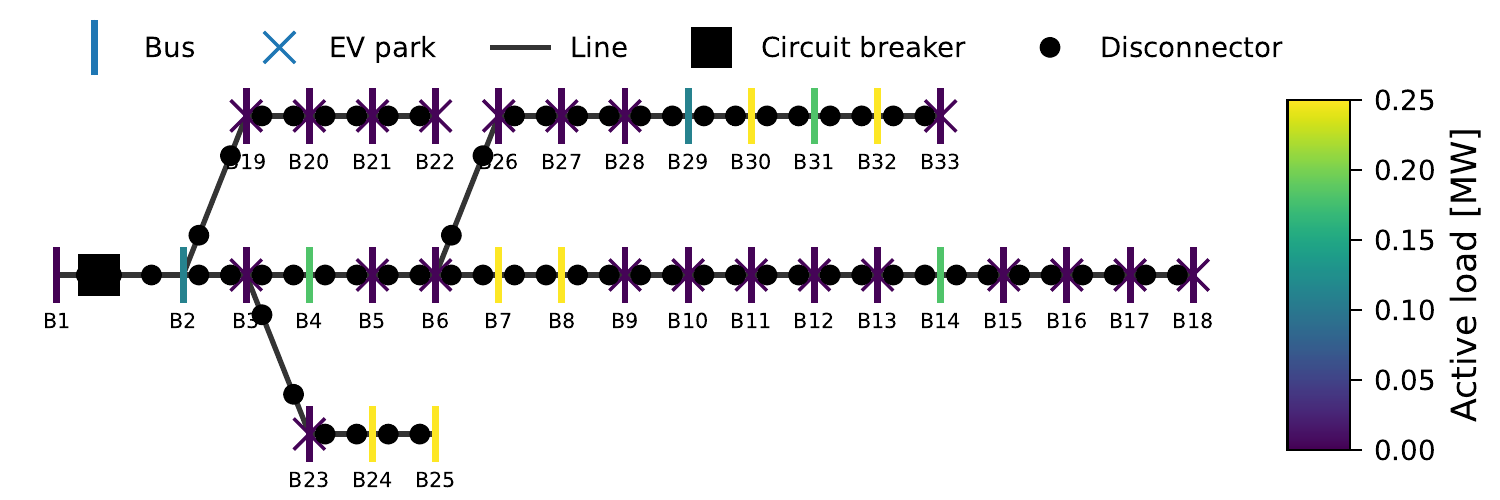}
  \caption{The system topology. The buses with EVs are shown with an X. The plot illustrates a heat map of the mean load at each bus in the system.}
  \label{fig:topology}
\end{figure*}

The method is demonstrated on the IEEE 33-bus network seen in Fig. \ref{fig:topology}. This network is chosen since it is a commonly used test network, making it easier for evaluation and replication. 
%The IEEE 33-bus distribution system is used to investigate the impact from V2G technology on the reliability of electricity supply in the distribution system. The network can be seen in Fig. \ref{fig:topology}. 
To account for a more realistic
distribution network, the customers in the network with
their demand profiles are distributed to map a real network. To make a dynamic demand profile of non-constant load, load profiles generated based on the FASIT requirement specification \cite{FASIT}. The generated load profiles give hourly time variations of the load. In order to evaluate the system with smaller time increments, the load is interpolated to the simulated time increment. The load from FASIT differs in customer groups and includes households, farms, trade, office buildings, and industry. The temperature data for generating the load profiles are collected from a weather station located in the east of Norway. In Fig.\ref{fig:topology}, the mean load at each bus in the network is illustrated through a heat map.

%The distribution system consists of different types of loads such as households, farms, trade, office buildings, and industry. The load profiles of the different load types are generated based on the FASIT requirement specification . The weather data is collected from a weather station located in the eastern Norway. %The reliability data for the outage time and the failure rate can be seen in Tab. \ref{tab:Reliability_data}. 
The repair time follows the distribution described in Sec. \ref{sec:outage_time_dist}. The repair time data is based on yearly reliability statistics from the Norwegian DSOs \cite{Outage_time}. The repair times used in this paper span over a time of up to two hours. This time is chosen based on two reason: 1) the availability of disruption data from the Norwegian DSOs \cite{Outage_time} and 2) this is a time span that will illustrate to what degree the EVs can support for some longer outage periods. 
Based on the reliability statistics from the Norwegian DSO, the failure rate corresponding the repair time statistics is 0.026 failures/year/km. This failure rate is based on a collection of the failure rate of multiple components with outage times up to two hours \cite{Outage_time}. 

\begin{comment}

\begin{table}[H]
    \caption{Reliability data of the distribution system}
    \center
\begin{tabular}{c|c} 
     \hline
     \textbf{Outage time [h]} & 0.5-2 \\ 
     \textbf{Failure rate [failure/year]} & 0.013 \\ 
     \hline
\end{tabular}
\label{tab:Reliability_data}
\end{table}

\end{comment}

\subsubsection{EVs in the distribution system}

The EV parks in the system are connected to the buses containing household loads. In Fig. \ref{fig:topology}, the buses with an X have EV parks connected. The availability of the EVs in each EV park follows the procedure outlined in Sec. \ref{sec:method}. 

The mapping of EVs compared to houses in the distribution system is sat at 46\%. This is based on the predicted average share of EVs in 2030 in Norway \cite{TOI}. With an EV share of 46\%, the distribution network will have a maximum of 290 cars. Due to the availability based on the individual charging pattern, the number of available EVs will be fewer.

%on data from giving a total of X maximum \stine{legg til dette basert på antall hus i nettet} EVs in the distribution network. This gives the total EVs in the EV park that might be available for V2G services to the distribution system.

\subsection{Scenario descriptions}

The case study investigates four different cases: 

\begin{itemize}
    \item \textbf{Case 1: EV}---No V2G, no support during failures in the network. 
    \item \textbf{Case 2: V2G}---V2G is activated, and the available EVs can support during failures in the network.
    \item \textbf{Case 3: EV and battery}---V2G is not activated, batteries are included.
    \item \textbf{Case 4: V2G and battery}---V2G is activated and batteries are included. 
\end{itemize}

The cases aim to investigate the difference in impact by having V2G support compared to no support (only regular EVs). In Cases 1 and 2, the impact from V2G is addressed when no other generation sources are available. In addition, Cases 3 and 4 are included to investigate the impact of V2G when other energy sources are added. In Cases 3 and 4, two batteries are integrated into the system at B18 and B33. These batteries are implemented to serve as a backup supply for the network and will be fully charged. The reason for including the batteries is to pinpoint the actual impact V2G serves. Since the charging capacity and the battery capacity of EVs are low, there is a limited amount of energy the EVs can possibly support. By including other sources, the opportunity to identify the impact might increase. 

%the impact of V2G are investigated by also including other energy sources. 

The data used in the scenarios are summarized in Tab. \ref{tab:Scenario_data}. 

\begin{comment}
\begin{table}[H]
    \caption{Scenario data}
    \center
    \resizebox{0.5\textwidth}{!}{\begin{tabular}{|c||c||c|c|} 
     \hline
     \textbf{Charging capacity [kW]} & \textbf{Share of EVs [\%]} & \multicolumn{2}{c|}{\textbf{Outage time distribution}} \\ \hline 
     \multirow{2}{*}{3.6} & \multirow{2}{*}{0.46} & \multirow{2}{*}{Normal} & Loc: 1 \\
      & & & Scale: 0.5\\ \hline
\end{tabular}}
\label{tab:Scenario_data}
\end{table}

\end{comment}

\begin{table}[H]
    \caption{Data for the case study}
    \center
    \resizebox{0.5\textwidth}{!}{\begin{tabular}{|l||c|c|} 
     \hline
     \textbf{Parameter} & \multicolumn{2}{c|}{\textbf{Value}} \\ \hline
     EV battery capacity [kWh] & \multicolumn{2}{c|}{70} \\ \hline
     Charging capacity EVs [kW] & \multicolumn{2}{c|}{3.6} \\ \hline
     Share of EVs [\%] &\multicolumn{2}{c|}{0.46}  \\ \hline
     \multirow{2}{*}{Outage time distribution} & \multirow{2}{*}{Normal} & Loc: 1 \\ 
     & & Scale: 0.5 \\ \hline
     \hline
     Battery capacity [MWh] & \multicolumn{2}{c|}{0.5}  \\ \hline
     Inverter capacity [MW] & \multicolumn{2}{c|}{0.25}  \\ \hline
     Efficiency & \multicolumn{2}{c|}{0.95}  \\ \hline
     Min SoC & \multicolumn{2}{c|}{0.1}  \\ \hline
     
\end{tabular}}
\label{tab:Scenario_data}
\end{table}

\begin{comment}

\begin{table}[H]
    \caption{Scenario data}
    \center
\begin{tabular}{|c|c|c|} 
     \hline
     \textbf{Charging capacity [kW]} & 3.6  & \\ 
     \textbf{Share of EVs} & 0.46 &  \\ 
     \multirow{2}{*}{\textbf{Outage time distribution}} & \multirow{2}{*}{Normal distribution} & Loc: 1.25\\
     & & Scale: 0.1 \\  
     \hline
\end{tabular}
\label{tab:Scenario_data}
\end{table}

\end{comment}

\subsection{Sensitivity analysis}

In addition to the case study, % that investigates the effect of having V2G possibilities during outages in the distribution system, 
a sensitivity analysis is conducted to investigate the importance of the different parameters. The sensitivity analysis is conducted as a full factorial design where the effect independent input parameters have on both the output variables and the other input parameters. The investigated parameters are:

\begin{itemize}
    \item \textbf{Charging capacity}---Since the charging capacity on the EVs can vary, we will investigate which effect the invert capacity has on the reliability of the system and the EVs in the network.
    \item \textbf{Share of EVs in the network}---The share of EVs in the distribution network is increasing rapidly, and in the future, some locations are estimated to have a share of EVs close to 100\% EVs \cite{TOI}. Therefore, different percentages of EV share ($X_{EV}$) in the network will be investigated. This is also to investigate how an increased share of EVs will impact the reliability of the distribution network. 
    \item \textbf{Repair time}---The study investigates various repair times of up to two hours. To evaluate the effect of the repair time, the repair time distribution will be adjusted to investigate the reliability impact and the impact on the EVs in the system. 
\end{itemize}

The parameter values used in the factorial design can be seen in Tab. \ref{tab:Sensitivity_data}.

\begin{comment}
\begin{table}[H]
    \caption{Sensitivity analysis data}
    \center
\begin{tabular}{c|ccc} 
     \hline
     \textbf{Charging capacity [kW]} & 3.6 &  &  7.2 \\ 
     \textbf{Share of EVs } & 0.46 & 0.61 & 0.87 \\ 
     \textbf{Outage time distribution} & often low & normal distribution & often high \\ 
     \hline
\end{tabular}
\label{tab:Sensitivity_data}
\end{table}
\end{comment}

\begin{table}[H]
    \caption{Sensitivity analysis data}
    \center
    \resizebox{0.5\textwidth}{!}{\begin{tabular}{|c|c||c|c||c|c|c|} 
     \hline
     \multicolumn{2}{|c||}{\textbf{Charging capacity [kW]}} & \multicolumn{2}{|c||}{\textbf{EV share [\%]}} & \multicolumn{3}{|c|}{\textbf{Outage time distribution}} \\ \hline
     \multirow{2}{*}{$\mathbf{P_{1}}$} & \multirow{2}{*}{3.6} & \multirow{2}{*}{$\mathbf{X_{1}}$} &\multirow{2}{*}{0.46} & \multirow{2}{*}{$\mathbf{r_{1}}$} & \multirow{2}{*}{Low} & Loc: 0.5 \\
      & & & & &  & Scale: 0.5 \\ \hline
      \multicolumn{2}{|c||}{} & \multirow{2}{*}{$\mathbf{X_{2}}$} & \multirow{2}{*}{0.61} & \multirow{2}{*}{$\mathbf{r_{2}}$}& \multirow{2}{*}{Normal} & Loc: 1 \\
      \multicolumn{2}{|c||}{} & & & &  & Scale: 0.5 \\ \hline
      \multirow{2}{*}{$\mathbf{P_{2}}$}  & \multirow{2}{*}{7.2} & \multirow{2}{*}{$\mathbf{X_{3}}$}  &\multirow{2}{*}{0.87} & \multirow{2}{*}{$\mathbf{r_{3}}$} & \multirow{2}{*}{High} & Loc: 1.5 \\
      & & & & &  & Scale: 0.5 \\  
     \hline
\end{tabular}}
\label{tab:Sensitivity_data}
\end{table}
\section{Result and discussion}
\label{sec:result}

This section aims to present the results from the case study and the sensitivity analysis. The results from the case study are detailed first before the results from the sensitivity analysis are given. Finally, a discussion of the result is presented. 

Each simulation in the study is simulated with 3,000 iterations where the same random seed is used for all cases to ensure the same basis for all the cases. The convergence is achieved after approximately 1,500 iterations. The increment value is of five minutes. The example network with the dataset can be located in the documentation page of the software \cite{RELSAD_documentation}. 

\subsection{Scenario results}

%In Fig. \ref{fig:box_plot_EV_V2G}, a box plot illustrating the frequency distribution of $\tt ENS$ in the distribution system for the four cases, is presented. There can be observed a decrease in ENS when V2G service is provided to the distribution system. The average ENS in the distribution system for both cases can be seen in Tab. \ref{tab:ReliabilityIndices_system}. The decrease in ENS is on 5.51\% giving a statistical significant improvement. 

\begin{figure}[h]
    \centering
    \includegraphics[width=0.5\textwidth]{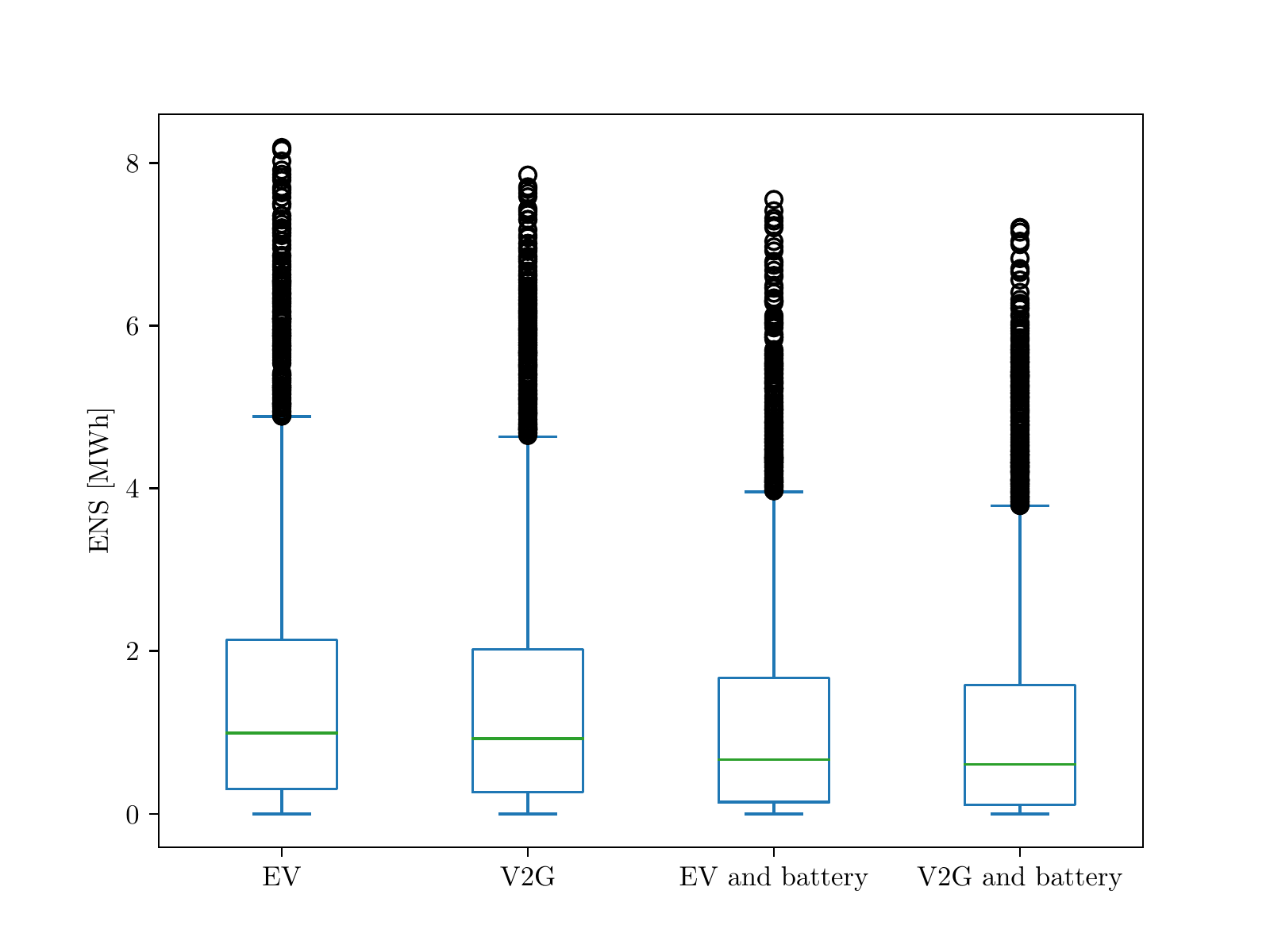}
  \caption{Box plot of the frequency distribution of ENS in the distribution system.}
  \label{fig:box_plot_EV_V2G}
\end{figure}

In Fig. \ref{fig:box_plot_EV_V2G}, a box plot illustrating the frequency distribution of $\tt ENS$ in the distribution system for the four cases is presented. 
%Fig. \ref{fig:bar_plot_all} gives the mean values for all the indices for the simulated cases. 
We observe a decrease in $\tt ENS$ when V2G service is provided to the distribution system, compared to their respective cases without V2G services. The average $\tt ENS$ in the distribution system for the cases can be seen in Tab. \ref{tab:ReliabilityIndices_system}. The decrease in $\tt ENS$ is 5.51\% when comparing Case 1 and Case 2. By comparing Case 3 and Case 4, the decrease in $\tt ENS$ is 6.39\% when activating V2G. The results indicate that, with V2G, the EVs are able to support the system for short outages. This is also shown in Case 4 where the battery can provide for longer outages and the EVs can serve as a support to the battery.

By investigating $\tt SAIFI$ and $\tt SAIDI$ for the cases, seen in Tab. \ref{tab:ReliabilityIndices_system}, $\tt SAIFI$ is decreasing between Case 1 and Case 2 and between Case 3 and Case 4. This illustrates that the total number of interruptions decreases by activating V2G. However, by evaluating $\tt SAIDI$, we see that the decrease is not significant. This indicates that the load shedding is due to the reduction in the number of interruptions the load points experience but that the interruption is prevented for short outages only.

The results for the EV indices are also shown in Tab. \ref{tab:ReliabilityIndices_system}. As expected, the average EV demand not served for the EVs will almost double when V2G is activated. This is a result of all the available EVs being used for V2G services and the demand not served increases since the EVs will be discharged as well. 
The other two EV indices, $\tt EV_{Int}$ and $\tt EV_{Dur}$, will be zero for the cases without V2G since the EVs will not be used as a service. For the two cases with V2G activated, it is evident that on average an EV owner can expect to experience the EV being used for support 0.15 times during a year for Case 1. This is a little higher for Case 4 when the batteries are introduced, where the expected interruption is then 0.32 times a year. The reason for this increase could be that when the batteries are used as the main provider and the EVs are there for support, there could be time increments where the battery can provide the necessary demand but others where the EVs need to be used. This will result in the V2G support being turned on and off during an outage and the total number of interruptions will increase. 
In both Case 2 and Case 4, the EVs are supporting an average of 38 minutes during a year. These results illustrate that the EVs are neither used often nor for long durations during a year when these reliability parameters are considered.

 %Tab. \ref{tab:ReliabilityIndices_system} also illustrates the other reliability indices for the system. Both the average experienced interruption frequency and duration for the system decreases. SAIFI is almost halved while SAIDI has decreased from 1.29 hours to 1 hour. CAIDI, however, will increase when V2G services are applied. Since CAIDI is a production of the fraction of SAIDI and SAIFI, 

%The EV oriented indices are presented in Tab. \ref{tab:ReliabilityIndices_EV}. The results show that the amount of power the EVs lose is close to doubled when V2G services are applied in the system. The result is expected since the EVs will in addition to not being able to charge, discharge power to the system. For such large distribution systems and the low discharging capacity of the EVs, all the available EVs will be used for support. 

%From the other two EV indices, it can be seen that the EVs on average are used with a frequency of 1.2 times a year with average support duration of 30 minutes. 

\begin{comment}
\begin{figure}[h]
    \centering
    \includegraphics[width=0.5\textwidth]{Figures/indices_mean_bar_plot.pdf}
  \caption{Interaction plot of the ENS for the distribution system}
  \label{fig:bar_plot_all}
\end{figure}
\end{comment}

%V2G reduserer total lastshedding med litt over 5 prosent. Basert på SAIFI indeksene virker det som at denne reduksjonen i lastshedding skyldes en reduksjon i antall avbrudd. Dette kan indikere at elbilene klarer å holde systemet gående for korte avbrudd.

\begin{table}[h]
    \caption{Distribution system and EV oriented reliability indices}
    \center
\begin{tabular}{|l|cccc|} 
     \hline
     & \textbf{Case 1} & \textbf{Case 2} & \textbf{Case 3} & \textbf{Case 4}\\ \hline
     $\tt ENS $ [MWh] & 1.5809 & 1.4938 & 1.2529 & 1.1732\\ 
     $\tt SAIFI $ [-] & 0.4011 & 0.3881 & 0.3392 & 0.3142\\ 
     $\tt SAIDI $ [h] & 1.1348 & 1.1346 & 1.0200 & 0.9998\\ 
     \hline \hline 
     $\tt EV_{Demand}$ [MWh] & 0.0362 & 0.0708 & 0.0055 & 0.0116\\ 
     $\tt EV_{Dur}$ [h] & 0.00 & 0.6336 & 0.00 & 0.6461 \\ 
     $\tt EV_{Int}$ [-] & 0.00 & 0.1546 & 0.00 & 0.3185 \\ 
     \hline 
\end{tabular}
\label{tab:ReliabilityIndices_system}
\end{table}

\begin{comment}

\begin{table}[H]
    \caption{EV oriented reliability indices}
    \center
\begin{tabular}{l|cc} 
     \hline
     & \textbf{Case: EV} & \textbf{Case: V2G} \\ \hline
     \textbf{$\tt EV_{Index}$ [MWh]} & 0.0403 & 0.0777 \\ 
     \textbf{$\tt EV_{Int}$ [-]} & 0.00 & 1.2305 \\ 
     \textbf{$\tt EV_{Dur}$ [h]} & 0.00 & 0.2794 \\ 
     \hline
\end{tabular}
\label{tab:ReliabilityIndices_EV}
\end{table}

\end{comment}

\subsection{Sensitivity analysis}

\begin{figure}[h]
    \centering
    \includegraphics[width=0.5\textwidth]{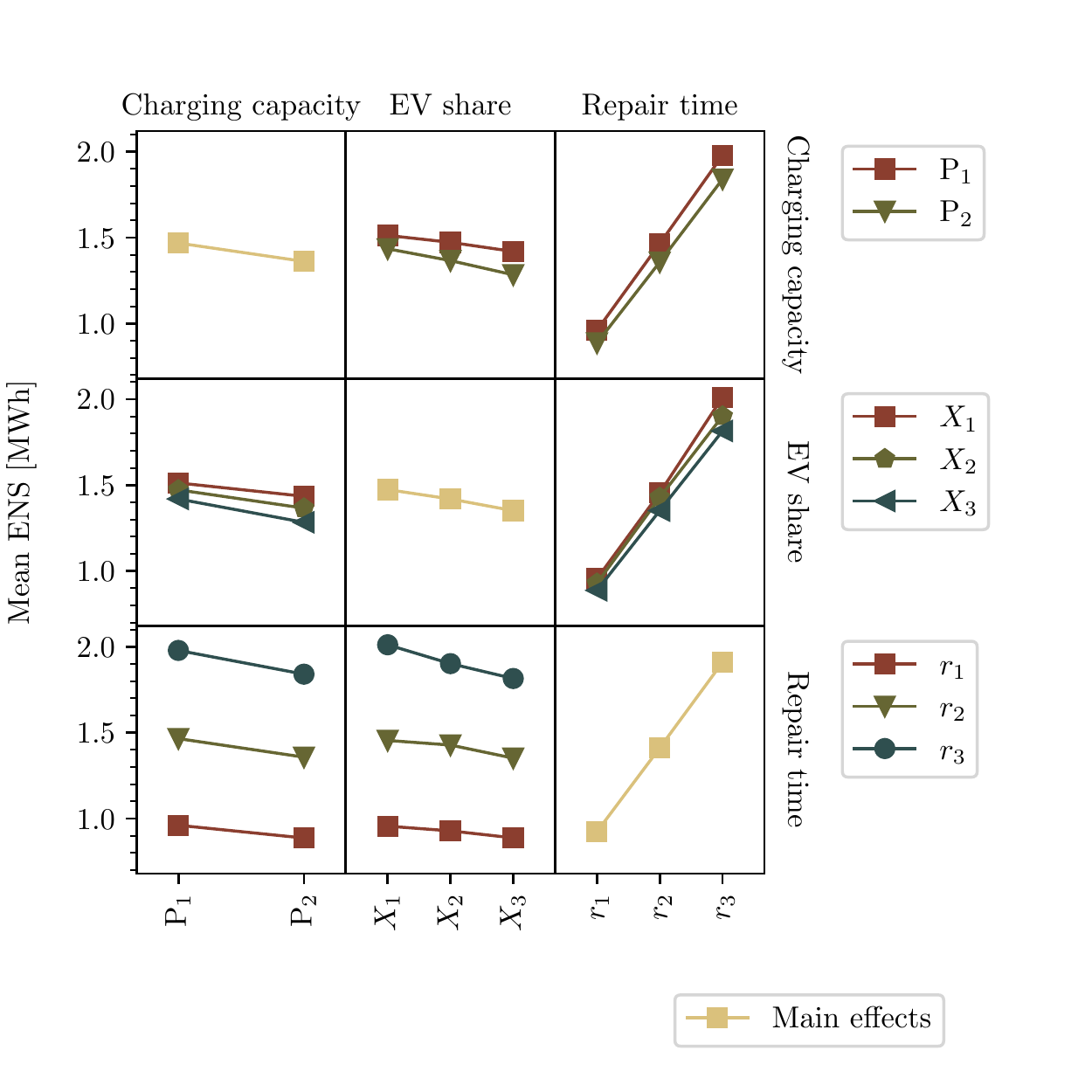}
  \caption{Interaction plot of the mean $\tt ENS$ of the distribution system for the parameter's charging capacity, EV share, and repair time. The plot illustrates that the repair time affects $\tt ENS$ where there is a small interaction effect between the charging capacity and the EV share.}
  \label{fig:interaction_plot_ENS}
\end{figure}

A full factorial design was conducted based on the parameters presented in Fig. \ref{tab:Sensitivity_data}. The sensitivity analysis was conducted on Case 2, and the results for all the presented indices are given. 
First, in Fig. \ref{fig:interaction_plot_ENS}, the interaction plot for mean $\tt ENS$ is shown. The figure shows that the repair time is the parameter that impacts the $\tt ENS$ the most. In addition, we observe a small interaction effect between the charging capacity and the share of EVs. This is an expected result, since a higher share of EVs in addition to more charging capacity results in more support. 

\begin{figure}[h]
    \centering
    \includegraphics[width=0.5\textwidth]{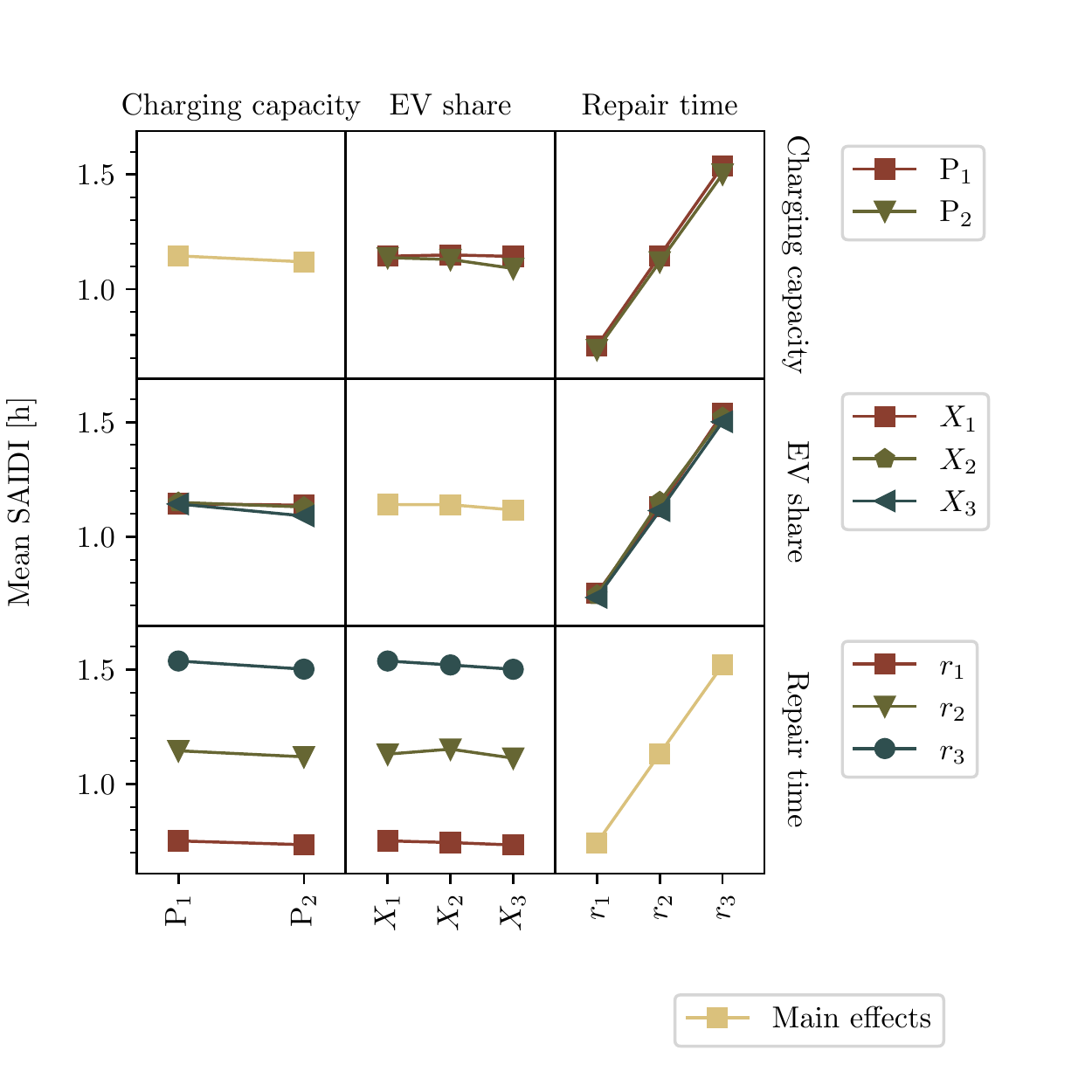}
  \caption{Interaction plot of the mean $\tt SAIDI$ indices of the distribution system for the parameter's charging capacity, EV share, and repair time. The plot illustrates that the repair time affects $\tt SAIDI$ whereas there is a small interaction effect between the charging capacity and the EV share.}
  \label{fig:interaction_plot_SAIDI}
\end{figure}

The interaction plots for mean $\tt SAIDI$ can be observed in Fig. \ref{fig:interaction_plot_SAIDI}. The interaction plot for $\tt SAIDI$ is very similar to the interaction plot for $\tt ENS$. The repair time gives the largest contribution as expected since a higher repair time gives longer down times in the system. The charging capacity and the share of EVs have almost no effect on $\tt SAIDI$, which was also illustrated in the results presented in Fig. \ref{tab:ReliabilityIndices_system}. Similar to the result for $\tt ENS$, we observe a weak interaction effect between the charging capacity and the share of EVs. This result indicates that higher EV share and charging capacity decrease the total down time and that might indicate that longer outages can be served. 

\begin{figure}[h]
    \centering
    \includegraphics[width=0.5\textwidth]{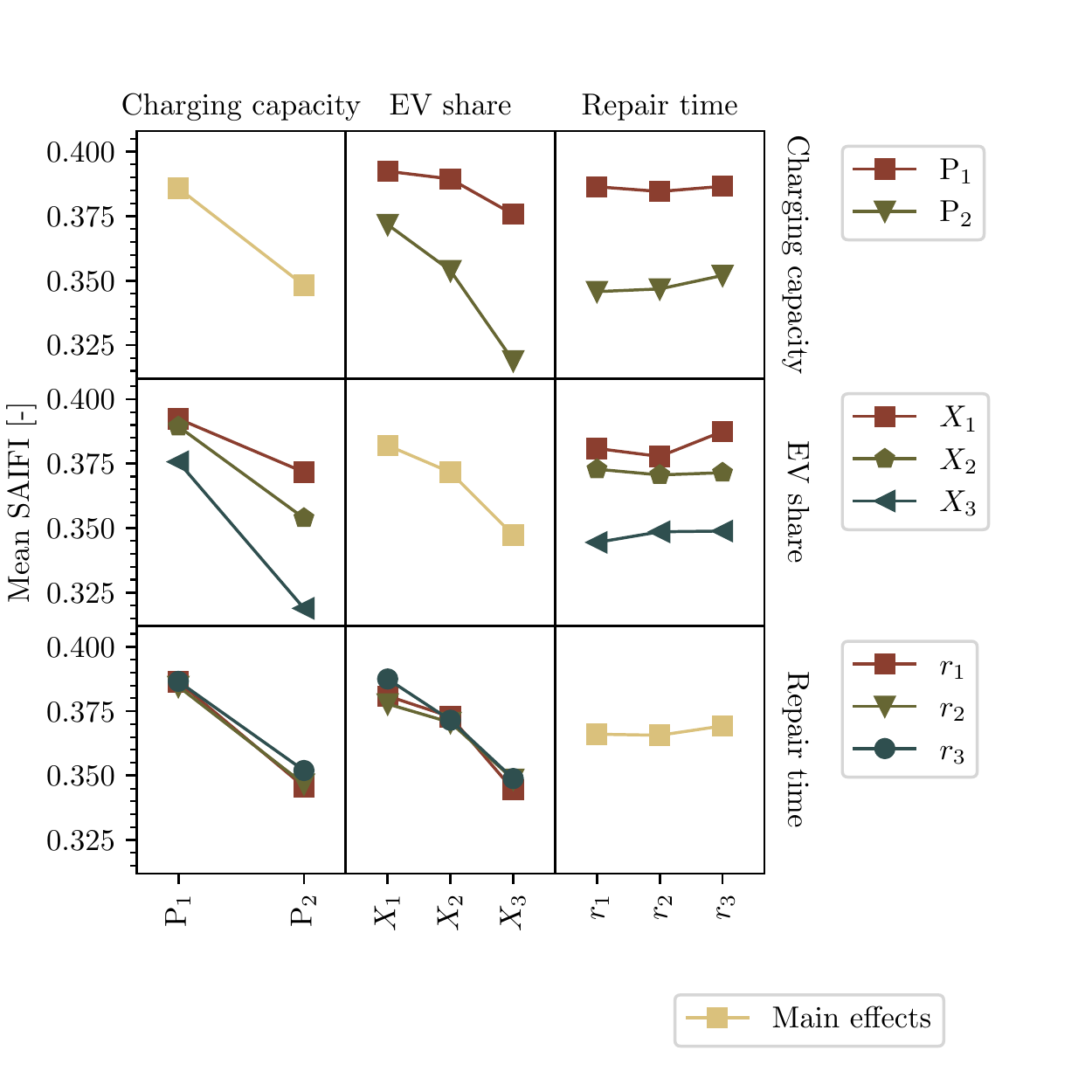}
  \caption{Interaction plot of the mean $\tt SAIFI$ indices of the distribution system for the parameters charging capacity, EV share, and repair time. The plot illustrates that the charging capacity and the EV share have a strong effect on the $\tt SAIFI$.}
  \label{fig:interaction_plot_SAIFI}
\end{figure}

For mean $\tt SAIFI$ seen in Fig. \ref{fig:interaction_plot_SAIFI}, the charging capacity and the share of EVs have a strong effect. Again, an interaction effect is seen between these two parameters. The interaction effect is stronger for $\tt SAIFI$ compared to the results for $\tt ENS$ and $\tt SAIDI$, which is a result of these parameters affecting the indices more. This is expected since a higher share of EVs and greater charging capacity enable the EVs to supply more load. The repair time, however, has a low effect and is only increasing the interruption frequency slightly for higher repair times. 

\begin{figure}[h]
    \centering
    \includegraphics[width=0.5\textwidth]{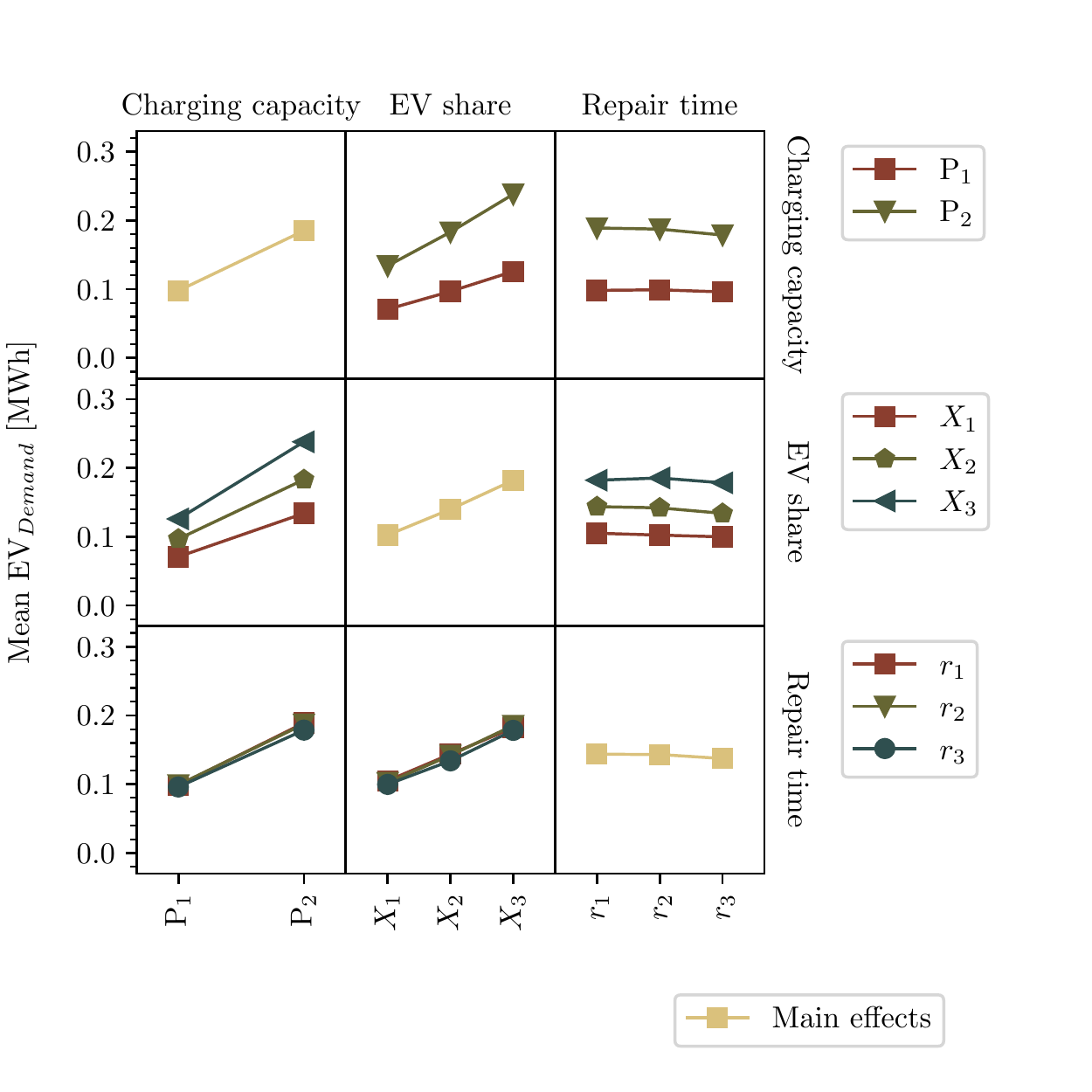}
  \caption{Interaction plot of the mean $\tt EV_{Demand}$ indices of the distribution system for the parameter's charging capacity, EV share, and repair time. The plot illustrates that the $\tt EV_{Demand}$ is affected by the charging capacity and the EV share.}
  \label{fig:interaction_plot_index}
\end{figure}

The interaction plot for the mean $\tt EV_{Demand}$ is presented in Fig. \ref{fig:interaction_plot_index}. The result illustrates that the demand for EVs is slightly affected by the repair time. The trend is decreasing, which could be a result of the EVs being emptied since the repair time is longer. The charging time, however, experiences a considerable increase in demand. With higher charging capacity, more power can be discharged from the EVs, and this will result in an increased demand for the EVs. The demand increases with an increased share of EVs as more EVs are in need of energy. As seen in the other results, there is a weak interaction effect between the charging capacity and the share of EVs for $\tt EV_{Demand}$ indices as well. 

\begin{figure}[h]
    \centering
    \includegraphics[width=0.5\textwidth]{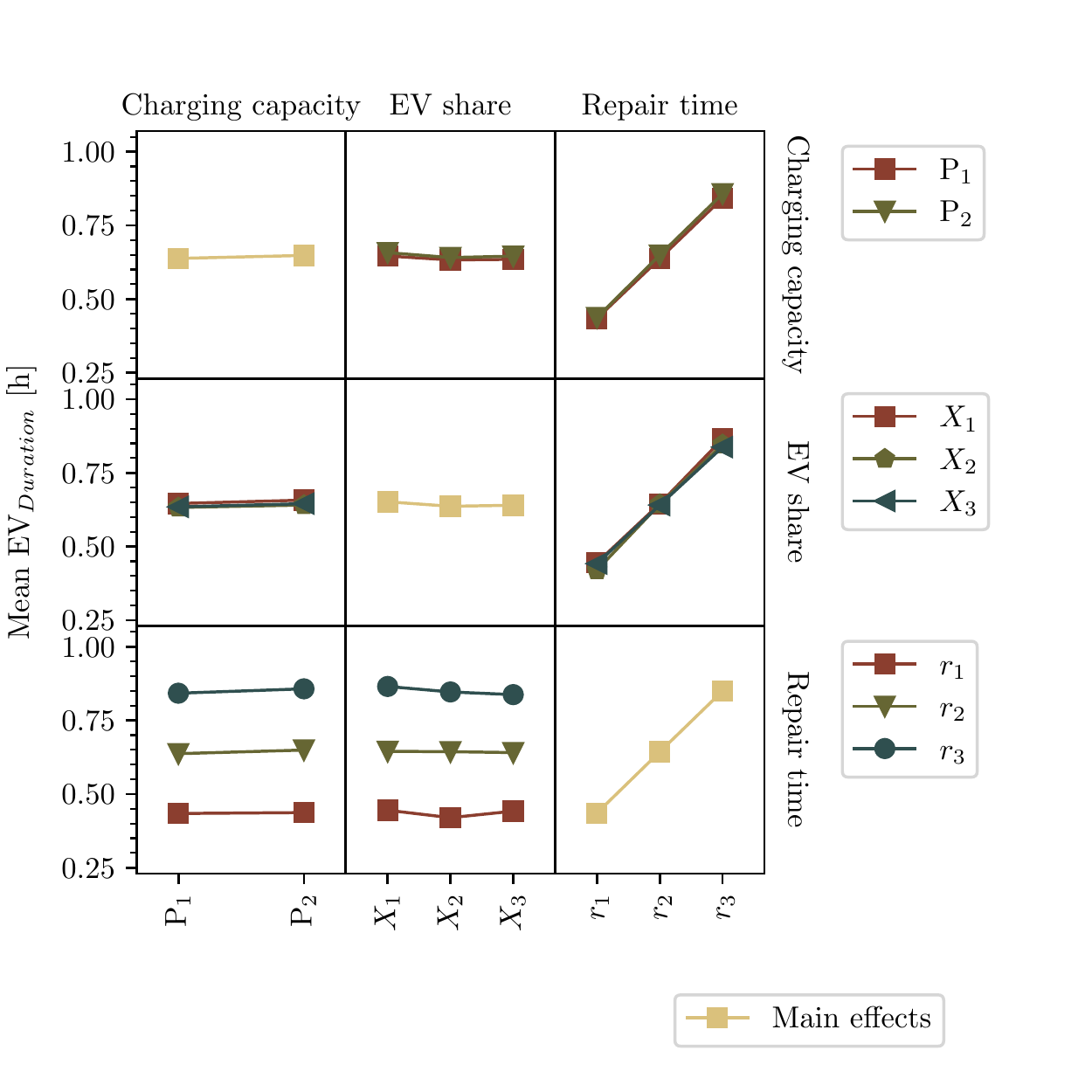}
  \caption{Interaction plot of the mean $\tt EV_{Dur}$ indices of the distribution system for the parameter's charging capacity, EV share, and repair time. The plot illustrates that the repair time affects the $\tt EV_{Dur}$.}
  \label{fig:interaction_plot_dur}
\end{figure}

The result for mean $\tt EV_{Dur}$ in Fig. \ref{fig:interaction_plot_dur} is similar to the results for mean $\tt ENS$ and $\tt SAIDI$. This is expected since this index is similar to $\tt SAIDI$. The charging capacity and the share of EVs have only small effects on the index. However, the repair time has a large effect since a longer repair time results in longer periods of support for the EVs.

\begin{figure}[h]
    \centering
    \includegraphics[width=0.5\textwidth]{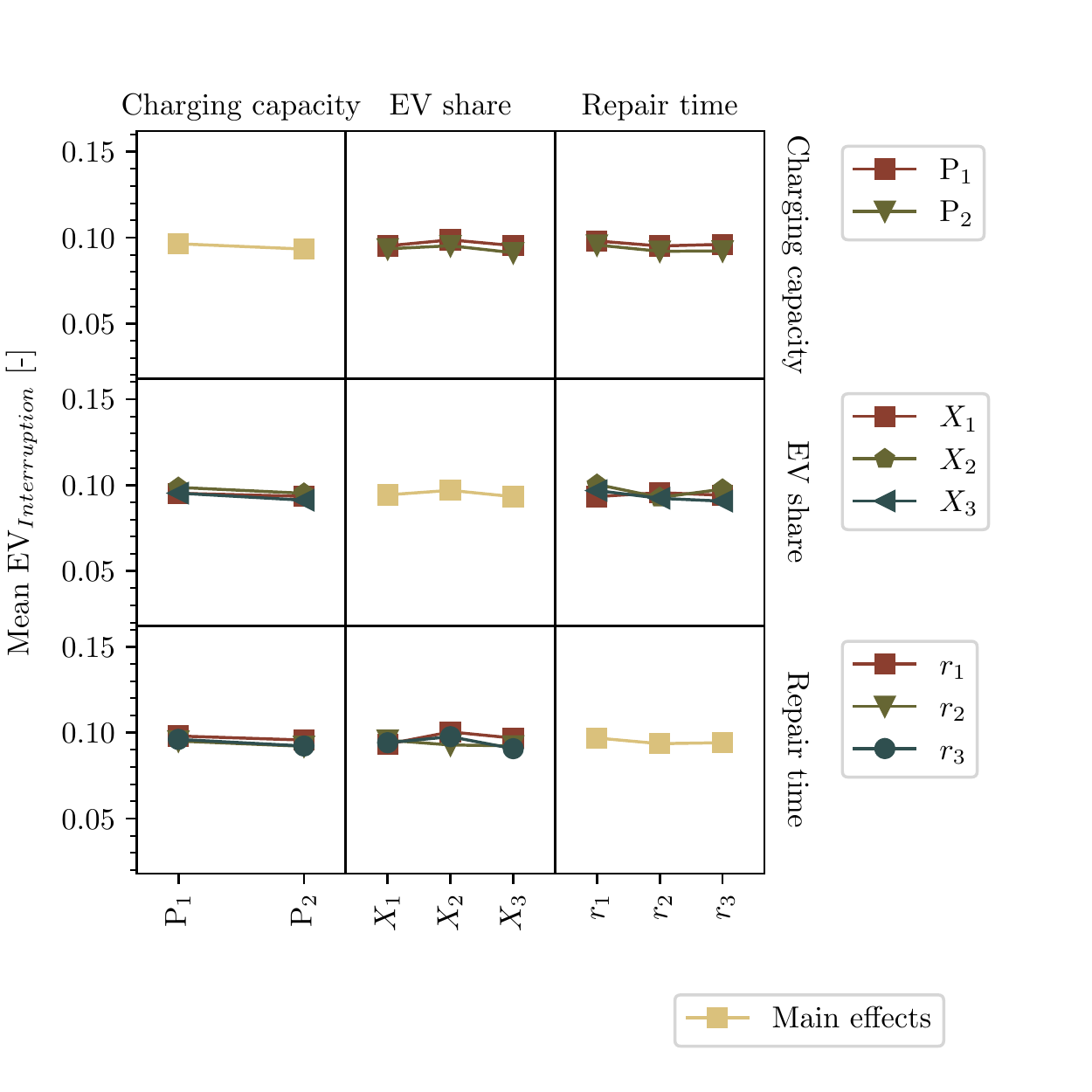}
  \caption{Interaction plot of the $\tt EV_{Int}$ indices of the distribution system for the parameter's charging capacity, EV share, and repair time. The results illustrate that $\tt EV_{Int}$ is not affected by any of the parameters.}
  \label{fig:interaction_plot_int}
\end{figure}

The interaction plot for mean $\tt EV_{Int}$ is illustrated in \cref{fig:interaction_plot_int}. We can see from the results that none of the parameters have any significant effect on the interruption of EVs in the system. We expect that the number of interruptions experienced by an EV will stay fairly constant until the distribution network is saturated. 

%The interaction plot for mean $\tt EV_{Int}$ is illustrated in Fig. \ref{fig:interaction_plot_int}. The total share of EVs affects this index the most. It seems that with a higher share of EVs, the fraction of EVs in an EV park compared to the used EVs decrease. The experience interruption is also decreasing with increased charging capacity, especially with higher EV share. The EVs can support more demand in a time step and with more available EVs, not all the EVs in an EV park are necessarily interrupted. However, an important note is that the changes in interruption are very small in this result compared to the result for the other indices. This is important to consider when investigating the result, whereas, for example, the result for the repair time is slightly decreasing for the middle value of the outage time. This can be a result of the variations in the simulations because of randomness. However, the trend shows that it is not very affected. 

%Liten inteaksjon mellom charging capacity og repair time. Hvorfor omvendt for repair time 2 med avaialbility 2 i forhold til de andre??? 

\subsection{Discussion}

When analyzing the results from the case study, it can be observed that the EVs do in fact have an impact on the reliability of the distribution network when V2G is activated. However, the cases indicated that this type of V2G support is more suitable for short outage periods. The effect from the V2G could be more significant if other generation source solutions are also available in the system, such as renewable energy sources and batteries. An indication of this was given in the case where V2G and batteries were included when longer outages could be provided for and the EVs served as a backup. This result seems to be a general result due to the small contributions the EVs can provide. Since the EVs are restricted by the charging capacity, the EVs are limited to small power contributions compared to other sources such as a battery. 
Since the simulated failure types have a low probability of occurring, the results indicate that the EVs are not used very often during a year.
Since the interruption frequency is so low, an EV owner will most likely not be affected within a year. The total duration an EV is used is also very low, under 40 minutes, meaning that most of the used EVs can restore their demand relatively fast after being used for support. This is encouraging for motivating participation in such V2G programs. The total degradation of the EV battery will also be very low when the EV is used so rarely and during short periods where the charging capacity is low. If EV owners also receive benefits for participating, increased participation can be expected. In addition, the benefits from V2G need to be assessed through a cost-benefit analysis.

%This was proved when batteries were implemented where The EVs could then help restore supply in the network as a backup solution to the larger available generation sources and for severe outage situations in the network.  

The duration-based indices ($\tt ENS$, $\tt SAIDI$, and $\tt EV_{Dur}$) are very sensitive to the repair time of the system components where the other parameters have small effects. The frequency-based ($\tt SAIFI$ and $\tt EV_{Int}$), however, are not. Longer repair time means a longer down time of the network, but for $\tt SAIFI$ and $\tt EV_{Int}$, longer repair time does not impact an already started interruption significantly. The share of EVs, however, will affect the frequency-based indices since more EVs mean that more load can be supplied and fewer EVs need to be used. However, this is not the case for $\tt EV_{int}$ since there is no saturation in the network and all the available EVs are used to support the network.  
The interaction effect between the charging capacity and the share of EVs is observed for multiple indices. This indicates that with higher charging capacity and a number of EVs, more support can be given. Something to note is that the results might be sensitive to the strategy for choosing which EV parks and EVs to use for the V2G service. In this study, the EV parks were ordered by the id of the parent bus and the EVs within the EV parks were chosen randomly.
%The EVs will then need to support for a longer duration and isolated buses will be without supply longer. However, the frequency based indices ($\tt SAIFI$ and $\tt EV_{Int}$) gives another result. 

\section{Conclusion}
\label{sec:conclusion}

In this study, the impact of V2G on the reliability of a modern distribution system for short repair times was investigated. The results illustrate that EVs used for V2G support do have an impact on the reliability. However, the support is more suitable for short down times and in combination with other generation sources where the EVs can serve as a backup. In addition, the impact on the EVs is minor since the cars are affected rarely for a short duration at the time. We provided three new EV-related indices that successfully captured the impact experienced by the EVs when V2G is used. The EV-related indices were used to gain valuable insight into the EV perspective of the V2G service. A thorough sensitivity analysis was conducted to analyze how the EV charging capacity, EV share, and repair time of components in the system affect the reliability indices of the system and the EVs. The EV indices give an estimate of both the frequency and duration the EVs are used for V2G services which further provides the impact the EVs will experience. The proposed indices seem to capture the overall system effect experienced by the EVs, and may therefore be suitable for use in further investigations. The presented methodology for evaluating how EVs perform from a reliability perspective using RELSAD shows great promise. The study investigated the reliability of electricity supply both from the perspective of the distribution network and from the perspective of the EVs. The tool facilitates analysis of multiple different cases and the results are detailed distributions that serve as a good basis for further analysis.

\section{Acknowledgment}
\label{Acknowledment}

This work is funded by CINELDI - Centre for intelligent electricity distribution, an 8-year Research Centre under the FME-scheme (Centre for Environment-friendly Energy Research, 257626/E20). The authors gratefully acknowledge the financial support from the Research Council of Norway and the CINELDI partners.

\bibliographystyle{IEEEtran}
\bibliography{References}

\begin{thebibliography}{10}
\providecommand{\url}[1]{#1}
\csname url@samestyle\endcsname
\providecommand{\newblock}{\relax}
\providecommand{\bibinfo}[2]{#2}
\providecommand{\BIBentrySTDinterwordspacing}{\spaceskip=0pt\relax}
\providecommand{\BIBentryALTinterwordstretchfactor}{4}
\providecommand{\BIBentryALTinterwordspacing}{\spaceskip=\fontdimen2\font plus
\BIBentryALTinterwordstretchfactor\fontdimen3\font minus
  \fontdimen4\font\relax}
\providecommand{\BIBforeignlanguage}[2]{{%
\expandafter\ifx\csname l@#1\endcsname\relax
\typeout{** WARNING: IEEEtran.bst: No hyphenation pattern has been}%
\typeout{** loaded for the language `#1'. Using the pattern for}%
\typeout{** the default language instead.}%
\else
\language=\csname l@#1\endcsname
\fi
#2}}
\providecommand{\BIBdecl}{\relax}
\BIBdecl

\bibitem{EU_target}
\BIBentryALTinterwordspacing
{European Union}, ``Regulation ({EU}) 2021/1119: European climate law,'' Tech.
  Rep., Jul 2021. [Online]. Available:
  \url{https://eur-lex.europa.eu/legal-content/EN/TXT/PDF/?uri=CELEX:32021R1119&from=EN}
\BIBentrySTDinterwordspacing

\bibitem{share_EU}
\BIBentryALTinterwordspacing
eurostat, ``How are emissions of greenhouse gases by the eu evolving?''
  accessed: 01.02.22. [Online]. Available:
  \url{https://ec.europa.eu/eurostat/cache/infographs/energy/bloc-4a.html}
\BIBentrySTDinterwordspacing

\bibitem{share_Norway}
\BIBentryALTinterwordspacing
{Miljøstatus}, ``Norwegian emissions and uptake of greenhouse gases (in
  {N}orwegian),'' 2021, accessed: 01.02.22. [Online]. Available:
  \url{https://miljostatus.miljodirektoratet.no/tema/klima/norske-utslipp-av-klimagasser/}
\BIBentrySTDinterwordspacing

\bibitem{EV_UK}
\BIBentryALTinterwordspacing
{Department for Transport}, ``Government takes historic step towards net-zero
  with end of sale of new petrol and diesel cars by 2030,'' Nov 2020, accessed:
  01.02.22. [Online]. Available:
  \url{https://www.gov.uk/government/news/government-takes-historic-step-towards-net-zero-with-end-of-sale-of-new-petrol-and-diesel-cars-by-2030}
\BIBentrySTDinterwordspacing

\bibitem{EV_EU}
\BIBentryALTinterwordspacing
{European Council}, ``Fit for 55 package: Council reaches general approaches
  relating to emissions reductions and their social impacts,'' accessed:
  15.07.22. [Online]. Available:
  \url{https://www.consilium.europa.eu/en/press/press-releases/2022/06/29/fit-for-55-council-reaches-general-approaches-relating-to-emissions-reductions-and-removals-and-their-social-impacts/}
\BIBentrySTDinterwordspacing

\bibitem{NTP_EN}
\BIBentryALTinterwordspacing
{Norwegian Ministry of Transport}, ``National transport plan 2022-2033,'' Tech.
  Rep., 2021. [Online]. Available:
  \url{https://www.regjeringen.no/en/dokumenter/national-transport-plan-2022-2033/id2863430/}
\BIBentrySTDinterwordspacing

\bibitem{TOI}
L.~Fridstrøm, ``Electrifying the vehicle fleet: {P}roject for {N}orway
  2018-2025 (in {N}orwegian),'' {TØI}, Tech. Rep., 2019.

\bibitem{mazumder2020ev}
M.~Mazumder and S.~Debbarma, ``Ev charging stations with a provision of v2g and
  voltage support in a distribution network,'' \emph{IEEE Systems Journal},
  vol.~15, no.~1, pp. 662--671, 2020.

\bibitem{zecchino2019large}
A.~Zecchino, A.~M. Prostejovsky, C.~Ziras, and M.~Marinelli, ``Large-scale
  provision of frequency control via v2g: The bornholm power system case,''
  \emph{Electric Power Systems Research}, vol. 170, pp. 25--34, 2019.

\bibitem{bayati2019short}
M.~Bayati, M.~Abedi, G.~B. Gharehpetian, and M.~Farahmandrad, ``Short-term
  interaction between electric vehicles and microgrid in decentralized
  vehicle-to-grid control methods,'' \emph{Protection and Control of Modern
  Power Systems}, vol.~4, no.~1, pp. 1--11, 2019.

\bibitem{dinkhah2019v2g}
S.~Dinkhah, C.~A. Negri, M.~He, and S.~B. Bayne, ``V2g for reliable microgrid
  operations: Voltage/frequency regulation with virtual inertia emulation,'' in
  \emph{2019 IEEE Transportation Electrification Conference and Expo
  (ITEC)}.\hskip 1em plus 0.5em minus 0.4em\relax IEEE, 2019, pp. 1--6.

\bibitem{yumiki2022autonomous}
S.~Yumiki, Y.~Susuki, Y.~Oshikubo, Y.~Ota, R.~Masegi, A.~Kawashima,
  A.~Ishigame, S.~Inagaki, and T.~Suzuki, ``Autonomous vehicle-to-grid design
  for provision of frequency control ancillary service and distribution voltage
  regulation,'' \emph{Sustainable Energy, Grids and Networks}, vol.~30, p.
  100664, 2022.

\bibitem{masrur2017military}
M.~A. Masrur, A.~G. Skowronska, J.~Hancock, S.~W. Kolhoff, D.~Z. McGrew, J.~C.
  Vandiver, and J.~Gatherer, ``Military-based vehicle-to-grid and
  vehicle-to-vehicle microgrid—system architecture and implementation,''
  \emph{IEEE Transactions on Transportation Electrification}, vol.~4, no.~1,
  pp. 157--171, 2017.

\bibitem{farzin2016reliability}
H.~Farzin, M.~Fotuhi-Firuzabad, and M.~Moeini-Aghtaie, ``Reliability studies of
  modern distribution systems integrated with renewable generation and parking
  lots,'' \emph{IEEE Transactions on Sustainable Energy}, vol.~8, no.~1, pp.
  431--440, 2016.

\bibitem{xu2015reliability}
N.~Xu and C.~Chung, ``Reliability evaluation of distribution systems including
  vehicle-to-home and vehicle-to-grid,'' \emph{IEEE transactions on power
  systems}, vol.~31, no.~1, pp. 759--768, 2015.

\bibitem{Myhre2022}
\BIBentryALTinterwordspacing
S.~F. Myhre, O.~B. Fosso, P.~E. Heegaard, and O.~Gjerde, ``Relsad: A python
  package for reliability assessment of modern distribution systems,''
  \emph{Journal of Open Source Software}, vol.~7, no.~78, p. 4516, 2022.
  [Online]. Available: \url{https://doi.org/10.21105/joss.04516}
\BIBentrySTDinterwordspacing

\bibitem{RELSAD_documentation}
S.~F. Myhre, ``{RELSAD}-{R}eliability tool for {S}mart and {A}ctive
  {D}istribution systems,'' \url{https://relsad.readthedocs.io/en/latest/},
  2022.

\bibitem{IEA_EV}
\BIBentryALTinterwordspacing
{IEA}, ``Global {EV} {O}utlook 2021,'' Tech. Rep., 2021. [Online]. Available:
  \url{https://www.iea.org/reports/global-ev-outlook-2021}
\BIBentrySTDinterwordspacing

\bibitem{sorensen2021analysis}
{\AA}.~L. S{\o}rensen, K.~B. Lindberg, I.~Sartori, and I.~Andresen, ``Analysis
  of residential ev energy flexibility potential based on real-world charging
  reports and smart meter data,'' \emph{Energy and Buildings}, vol. 241, p.
  110923, 2021.

\bibitem{thingvad2018influence}
A.~Thingvad and M.~Marinelli, ``Influence of v2g frequency services and driving
  on electric vehicles battery degradation in the nordic countries,''
  \emph{Evs31}, 2018.

\bibitem{wang2016quantifying}
D.~Wang, J.~Coignard, T.~Zeng, C.~Zhang, and S.~Saxena, ``Quantifying electric
  vehicle battery degradation from driving vs. vehicle-to-grid services,''
  \emph{Journal of Power Sources}, vol. 332, pp. 193--203, 2016.

\bibitem{statnettibber}
\BIBentryALTinterwordspacing
{Statnett SF}, ``Distributed balancing of the power grid - results from the
  eflex pilot in the mfrr-market,'' Tech. Rep., 2021. [Online]. Available:
  \url{https://www.statnett.no/om-statnett/nyheter-og-pressemeldinger/nyhetsarkiv-2021/sikrer-stromforsyningen-med-bidrag-fra-elbiler-panelovner-og-ventilasjonsanlegg/}
\BIBentrySTDinterwordspacing

\bibitem{Virta}
\BIBentryALTinterwordspacing
Virta, ``Energy flexibility - how to unlock the real value of evs,'' accessed:
  01.02.22. [Online]. Available:
  \url{https://www.virta.global/blog/how-to-unlock-the-full-value-of-electric-vehicles-as-energy-flexibility-assets}
\BIBentrySTDinterwordspacing

\bibitem{Billinton}
R.~Billinton and R.~N. Allan, \emph{Reliability Evaluation of Power
  Systems}.\hskip 1em plus 0.5em minus 0.4em\relax New York: Springer
  Science+Business Media, LLC, 1996, no. 2nd ed.

\bibitem{elbilisten}
\BIBentryALTinterwordspacing
T.~{N}orwegian {E}lectric~{V}ehicle {A}ssociation, ``Norwegian {EV} owners
  survey (in {N}orwegian),'' 2017, accessed: 22.02.22. [Online]. Available:
  \url{https://elbil.no/elbilisten/}
\BIBentrySTDinterwordspacing

\bibitem{sadeghianpourhamami2018quantitive}
N.~Sadeghianpourhamami, N.~Refa, M.~Strobbe, and C.~Develder, ``Quantitive
  analysis of electric vehicle flexibility: A data-driven approach,''
  \emph{International Journal of Electrical Power \& Energy Systems}, vol.~95,
  pp. 451--462, 2018.

\bibitem{billinton_reliability_1994}
R.~Billinton and W.~Li, \emph{Reliability Assessment of Electric Power Systems
  Using {Monte Carlo} Methods}.\hskip 1em plus 0.5em minus 0.4em\relax Springer
  {US}, 1994.

\bibitem{haque1996load}
M.~Haque, ``Load flow solution of distribution systems with voltage dependent
  load models,'' \emph{Electric Power Systems Research}, vol.~36, no.~3, pp.
  151--156, 1996.

\bibitem{FASIT}
A.~O. Eggen, ``Fasit kravspesifikasjon,'' Sintef Energy AS., Tech. Rep., 2016.

\bibitem{Outage_time}
\BIBentryALTinterwordspacing
{Statnett SF}, ``Yearly statistics 2018 {(in Norwegian)},'' Tech. Rep., 2018.
  [Online]. Available:
  \url{https://www.statnett.no/for-aktorer-i-kraftbransjen/systemansvaret/leveringskvalitet/statistikk/}
\BIBentrySTDinterwordspacing

\end{thebibliography}

% References:

\end{document}